\documentclass[showpacs, prb,twocolumn,preprintnumbers , superscriptaddress, aps]{revtex4-2}

\usepackage{color}
\usepackage{amsmath,amssymb}
\usepackage{pifont}
\usepackage{amssymb}  % More symbols
\usepackage{bbold}
\usepackage{float}
\usepackage{subfloat}

\usepackage[caption=false]{subfig}
\usepackage{tikz}
\usepackage{makecell}
\usepackage{subfig}
\usepackage{pifont}   % Ding symbols
\usepackage{graphicx} % Include figure files
\usepackage{dcolumn}  % Align table columns on decimal point
\usepackage{bm}       % bold math
\usepackage{multirow} % Table functions
\usepackage{placeins}% For making floats not move around everywhere
\usepackage[colorlinks]{hyperref}
\usepackage{mathtools}

\newcommand{\vect}[1]{\mathbf{#1}}
\newcommand{\ket}[1]{\left|{#1}\right\rangle}
\newcommand{\bra}[1]{\left\langle{#1}\right|}

\usepackage{breqn}
\usepackage{mathrsfs}
\DeclareMathOperator{\Tr}{Tr}

\usepackage{multirow}

\begin{document}
\title{Fidelity of the Kitaev honeycomb model under a quench}

	\author{Wesley Roberts}
%	\email{roberts.w@northeastern.edu}
	\affiliation{Department of Physics$,$ Northeastern University$,$ Boston MA 02115$,$ USA}
	\author{Michael Vogl}
		\affiliation{Physics Department$,$
		King Fahd University
		of Petroleum $\&$ Minerals$,$
		Dhahran 31261$,$ Saudi Arabia}
	\author{Gregory A. Fiete}
%	\email{g.fiete@northeastern.edu}
	\affiliation{Department of Physics$,$ Northeastern University$,$ Boston MA 02115$,$ USA}
	\affiliation{Department of Physics$,$ Massachusetts Institute of Technology$,$ Cambridge MA$,$ USA}
	
	%Somehow commas were causing a compilation error I fixed this - not sure why

	\begin{abstract}
	    Motivated by rapid developments in the field of quantum computing and the increasingly diverse nature of qubits, we theoretically study the influence that quenched outside disturbances have in an intermediately long time limit. We consider localized imperfections, uniform fields, noise, and couplings to an environment which we study in a unified framework using a prototypical but idealized interacting quantum device - the Kitaev honeycomb model. Our study focuses on the quantum state robustness in response to an outside magnetic field, a magnetic bath, magnetic noise, magnetic impurities, and a noisy impurity. As indicators for quantum robustness, we use the Uhlmann fidelty of the ground state and excited spinon states after a quench. We find that the time dependence of the fidelity often depends crucially on whether the system is gapped. We find that in the gapped case the fidelity decays to a constant value under noiseless quenches, while in a gapless system it exhibits algebraic decay. In all other situations studied, such as coupling to a bath and noisy quenches, both gapped and gapless systems exhibit a universal form for the long-time fidelity,  $Ce^{-\alpha t}t^{-\beta}$, where the values of $C$, $\alpha$, and $\beta$ depend on physical parameters such as system size, disturbance strength, etc. Therefore, our work provides estimates for the intermediate-long time stability of a quantum device and it suggests under what conditions there appear the hallmarks of an orthogonality catastrophe in the time-dependence of the fidelity. Our work provides engineering guidelines for quantum devices in quench design and system size.
	    
	    %As external perturbations become increasingly negligible for increasingly good quantum computers our estimates provide insights that might be useful for the engineering of quantum devices. For instance we find that the impact of certain types of quenches can be minimized by choosing specific excited states to work with or by adjusting the size of a device.
	\end{abstract}
\maketitle
\section{Introduction}
\label{Intro}

Processing of quantum information requires coherence of the quantum states used to encode and transmit information \cite{Wilde,RevModPhys.89.041003,Preskill2018quantumcomputingin}. Losing coherence can be disastrous for the reliability of computational results \cite{chuang1995quantum,PhysRevLett.75.3788,https://doi.org/10.48550/arxiv.1203.5813,steane1998quantum,steane1999efficient,knill2005quantum}. As an extreme example, if between controlled computations a state  evolves into an orthogonal one all information necessary for further processing is lost. Indeed even less extreme reductions in the coherence of quantum states over time introduce error into the computation that are best avoided. This has led to large efforts being deployed to develop more fault tolerant devices as well as methods for error correction \cite{PhysRevA.52.R2493,PhysRevA.98.052348,Campbell_2017, Raussendorf_2007, Chao_2018, Egan_2021,PhysRevA.57.127,preskill1998reliable,freedman2003topological,Nayak_2008,sarma2015majorana,urbanek2021mitigating,ai2021exponential,steane1999efficient,knill2005quantum}. Therefore, it becomes clear that it is important to study coherence properties of systems that might serve as the basis for a quantum computer or other quantum devices - most importantly to understand their time dependence under environmental perturbations.

In an ideal scenario one would like to study the time resolved response of a quantum device at arbitrary times. However, often it is not possible to study the evolution - of even idealized quantum devices - on arbitrary time scales. Indeed, the time evolution operator is a time ordered exponential and therefore it is a complicated object that can be difficult to compute exactly except in a few cases \cite{giscard2015exact}. The time evolution operator becomes increasingly difficult to compute approximately as one studies longer time scales, which is evidenced by various approximate analytic \cite{vogl2019analog,dyson1949radiation,dyson1949s,dyson1952divergence,blanes2009magnus,magnus1954exponential,klarsfeld1989exponential,fernandez2004magnus,wilcox1967exponential,vogl2019flow} and computational schemes \cite{PhysRevLett.91.147902,paeckel2019time,vogl2019analog,White_2004,suzuki1993improved} that are known to eventually break down. Even state-of-the-art numerical schemes for strongly correlated systems such as tensor network methods require sophisticated approaches to deal with the increasing computational cost at long times that can be traced back to the enormous size of an interacting Hilbert space \cite{White_2004,paeckel2019time,haegeman2016unifying,alhambra2021locally}. For this reason one often restricts studies to time scales that are of the most experimental relevance. At the early stages of  research into quantum devices the external influence through noise, environmental coupling etc. on a system could be considered as relatively strong. In such an experimental regime it makes sense to study coherence measures over relatively short timescales because coherence will be lost relatively quickly and therefore long times are not of much interest in this regime. However, as technological advances are made that exclude outside disturbances to an increasing extent, the regime of interest is being pushed to ever longer timescales \cite{Vepsalainen_2020, Herbschleb_2019}. Thus, keeping pace with and anticipating further advances, in the present work we study coherence measures over long timescales using an asymptotic approach that is valid for a weak external influence \cite{VANKAMPEN1974215,Marino}. We note that a second motivation for studying the long-time regime is that it is also the regime in which universal behavior is expected to to emerge \cite{Silva}.

To observe the emergence of such universal behavior, we study the impact that various kinds of external disturbances can have on the coherence of a quantum state. Here, we distinguish two cases. The first case is one where disturbances can be modeled by an ordinary Hermitian and time-independent Hamiltonian such as in the examples of an external magnetic field or a magnetic impurity. Here, the magnetic impurity might be due to a deposited speck of dirt and the magnetic field due to stray magnetic fields that might appear in an experimental setup and are difficult to shield. The second case is disturbances that are best modeled in a Lindblad master equation approach \cite{manzano2020short,lindblad1976generators} since their most economical description involves non-pure density matrices. Here, examples include local or uniform couplings to an external heat bath and also noisy external local or uniform fields \cite{Marino,PhysRevLett.118.140403,lindblad1976generators,pearle2012simple} - both are effects that can appear due to insufficient or leaky shielding from an environment. We will find that the long-time coherence in all cases we study has the same functional form, $\propto e^{-\alpha t} t^{-\beta}$ - parameters $\alpha$ or $\beta$ can be zero for specific kinds of situations. This result is expected to be a relatively robust feature that does not depend on the details of a physical system such as the specific form of an excitation spectrum etc. Rather it is  universal in that it depends only on quite general features like dimensionality or whether the system is gapped or the  types of band crossings that occur etc. We stress that this exponential decay of the coherence in the presence of a weak disturbance has also been shown to occur in 1D models  \cite{Marino,PhysRevB.94.014310} and in chaotic systems \cite{Goussev_2008}. Here, unlike these results from the literature, we will focus on an interacting 2D spin system - the honeycomb Kitaev model \cite{Kitaev,Chen_2008}- which is also a prototypical spin liquid \cite{Balents_2010,FIETE2012845}.

There are various probes of the stability of a quantum system, most of which measure the distance of a reference state from a comparable state that was subjected to some modification, for example, by taking a reference state and then comparing it to an evolved state. For instance, one could slowly turn on a perturbation or subject the system to a pulse or various pulses. Here, we consider the conceptually simple case of a quench, in which a perturbation is turned on instantaneously and left on at all later times. Naturally, there are various kinds of distance measures between states with many of them induced by norms. Examples include a distance defined via a quantum metric from the field of quantum geometry \cite{Berry_1988,Provost_1980}, the Loschmidt echo \cite{PhysRevA.30.1610,PhysRevLett.124.160603,Goussev:2012,GORIN200633}, and its generalization to density operators, the Uhlmann fidelity \cite{UHLMANN1976273,doi:10.1080/09500349414552171}. 

Our work employs the Loschmidt echo and the Uhlmann fidelity to study the robustness of states in the Kitaev honeycomb model \cite{Kitaev,Chen_2008}. We focus on the Kitaev model because it is interesting not only for its status as an exactly solvable model and a protypical example of a 2D spin liquid, but also because of its relevance for robust quantum computation \cite{Nayak_2008}: its excitations are highly robust against external influences \cite{Kitaev, Nayak_2008}. The model is known to host anyonic excitations, which are important to the field of topological quantum computing, where they are proposed for use as qubits due to their inherent stability \cite{Chen_2010,KITAEV20032,Nayak_2008}.  We are therefore interested in the robustness of the spin liquid ground state, which is also important in the study of topological quantum computing since it is this ground state that can play host to the anyonic excitations that would be useful to employ as robust qubits in toplogical quantum computing. 

Insights about ground state robustness are an important area of study that can complement the inherent stability of its anyonic excitations. Indeed, the ground state of Kitaev materials \cite{TREBST20221} must be robust enough to survive the introduction of anyonic excitations. To gain such crucial insight into the robustness of the model's ground state, we study long-time coherence measures of the Kitaev ground state. We will focus on important but relatively rarely studied noisy quenches or sudden weak coupling to a heat bath. Such an approach can be used to model localized holes in magnetic shielding of a device or coupling to the environment.

To supplement our results we also consider the case of excited spinon states under various quenches. The study of the stability of spinon excited states in the presence of quenches allows us to gain some deeper appreciation for the relevance of ground state stability when it comes to excited states. Particularly, we find that the coherence properties of the ground state predominantly determine the coherence of such excited states - at least in the case of a noiseless quench. Generically, we can expect the ground state coherence at minimum to arise as a modulation factor for excited state coherence.

Our work is structured as follows. In Sec. \ref{sec:model} we  review important properties of the honeycomb Kitaev model, including its exact solution that, expressed in terms of Jordan-Wigner fermions, serves as a starting point for the present work. In Sec. \ref{Models} we introduce specific models that we consider for quenched external disturbances - such as environmental coupling, noisy drives, and static fields. Sec. \ref{sec:math} presents the mathematical methods that are used for studying the effect of these quenches. We define the coherence measures that are used in subsequent computations as well as the relevant approximation methods. In Sec. \ref{sec:results} we present long-time asymptotic expressions for the coherence measures. We stress that we observe a universal form for long-time coherence. We also highlight specific differences in the dependence of coherence on different physical situations such as type of quench, magnetic field strength, and system size.  These results are also summarized formally in Tab. \ref{Tab:res1} and might serve as a guideline in engineering of quantum devices.

Sec. \ref{sec:results} is concerned with the study of the model's excited states. For the present work we restricted ourselves to the study of occupied spinon modes. Our results stress the importance of ground state coherence - in many cases it is a good indicator for the coherence of excited states. However, we also find cases where many excitations lead to a more robust state, the coherence of which decays much more slowly than for the ground state, an interesting feature of potential relevance to applications. Section \ref{sec:outlook} discusses potential directions for future work and  Sec. \ref{sec:conclusion} summarizes and concludes our discussion.

\section{Physical model} \label{sec:model}

\begin{figure}[htp]
    \centering
    \includegraphics[width=8.5cm]{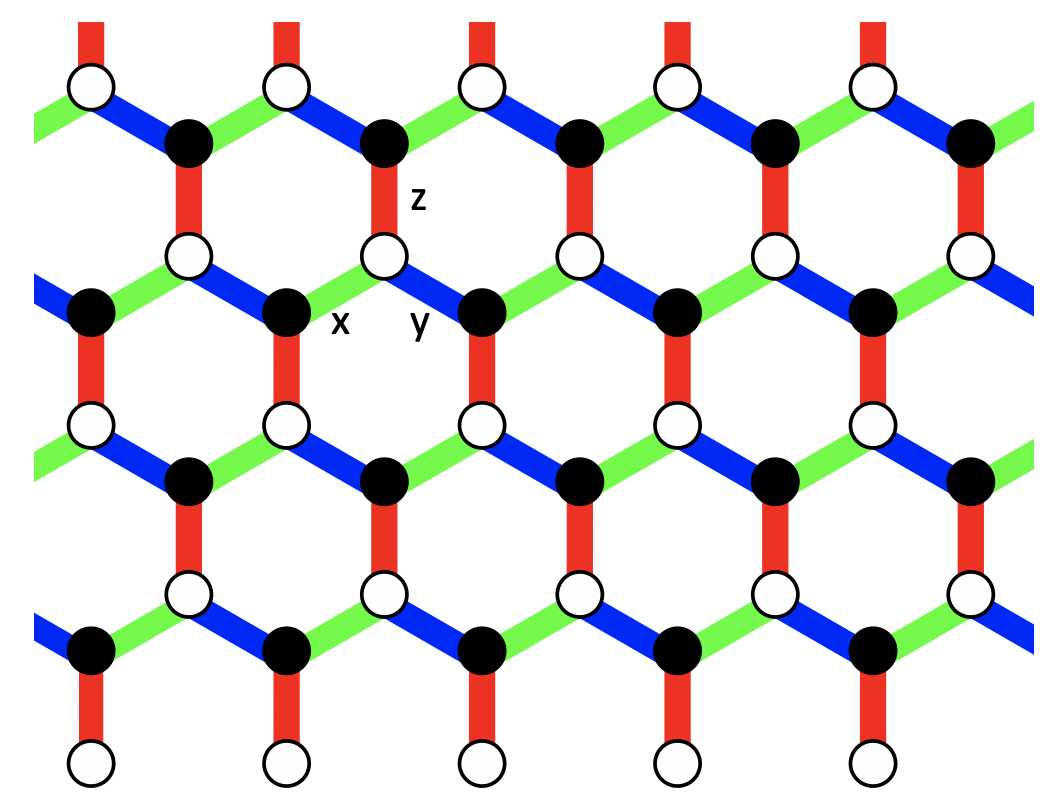}

    \caption{Kitaev honeycomb model, with couplings $J_x$, $J_y$, $J_z$ on the x-, y-, and z-bonds, respectively.  The different bond types are color coded.  \label{fig:honeycomb}}
    \end{figure}

We take the Kitaev honeycomb model \cite{Kitaev, Chen_2008} given in Eq.\eqref{eq:1} as a starting point. We begin with a summary of some of its equilibrium properties that will be needed for the discussion of a quenched model, which is the focus of our work. 

\begin{dmath} \label{eq:1}
    H = -J_x \sum_{x-bonds} \sigma^x_i \sigma^x_j - J_y \sum_{y-bonds} \sigma^y_i \sigma^y_j - J_z \sum_{z-bonds} \sigma^z_i \sigma^z_j.
\end{dmath}
The model describes a honeycomb lattice of moments with spin-$\frac{1}{2}$ at each site, with Pauli matrices $\sigma^x_i$, $\sigma^y_i$, and $\sigma^z_i$ for site $i$. Eq.\eqref{eq:1} can be diagonalized using a Jordan-Wigner transformation, which expresses the problem in terms of spinless fermions defined for each site \cite{WenQFT}. These Jordan-Wigner fermions can be subsequently represented in terms of fermions living on the $z$-bonds of the lattice, and another degree of freedom on $z$-bond $r$ represented by an operator $\alpha_r$, related to Kitaev's \cite{Kitaev} flux degree of freedom. For the details of this calculation the reader is referred to Ref.\cite{Chen_2008} or Appendix \ref{app:appendixA}. After these transformations the Hamiltonian becomes   

\begin{dmath} \label{Halpha}
H = \sum_r J_x(d^{\dagger}_r + d_r)(d^{\dagger}_{r+e_x} - d_{r+e_x})+ \sum_r J_y(d^{\dagger}_r + d_r)(d^{\dagger}_{r+e_y} - d_{r+e_y}) + \sum_r J_z\alpha_r(2d^{\dagger}_r d_r - 1),
\end{dmath}
where the sum is taken over $z$-bonds. We are interested in the ground state of the system, lying in the sector in which the operator $\alpha_r = 1$ for all $z$-bonds \cite{Chen_2008}. This reduces the problem to one of non-interacting fermions. The Hamiltonian is then readily diagonalized, giving (up to a constant offset energy)

\begin{equation} \label{eq:diagham}
    H = \sum_k E_k\gamma^{\dagger}_k \gamma_k
\end{equation}
where $E_k = \sqrt{\epsilon^2_k + \Delta^2_k}$ for $\epsilon_k = 2J_z -2J_x \cos{k_x} -  2J_y\cos{k_y}$ and $\Delta_k = 2J_x\sin{k_x} +2J_y\sin{k_y}$, with the spacing between $z$-bonds having been set to 1. We adopt this particular approach from \cite{Chen_2008} rather than the original method due to Kitaev \cite{Kitaev} because the latter involves introducing extra degrees of freedom, requiring projection into the physical Hilbert space from a larger artificial one. The method discussed in this section avoids technical complications that would arise in basing our calculations on Kitaev's original method.

The spectrum exhibits gapped and gapless phases depending on the parameters $J_x$, $J_y$, and $J_z$. When the parameters satisfy the triangle inequalities
$$J_x < J_y + J_z,$$
$$J_y < J_x + J_z,$$
$$J_z < J_x + J_y,$$
the spectrum is gapless \cite{Chen_2008}.
The quasiparticles $\gamma_k$ are spin excitations, and are related to the $d_k$ fermions via a Bogoliubov transformation
\begin{equation}
    d_k = u^*_k\gamma_k + v_k \gamma^{\dagger}_{-k},
\end{equation}
where $\left|u_k\right|^2 = \frac{1}{2}\left[1 + \frac{\epsilon_k}{E_k} \right]$ and $\left|v_k\right|^2 = \frac{1}{2}\left[1 - \frac{\epsilon_k}{E_k} \right]$.

The ground state can then be specified in terms of $d_k$ fermions using the condition that $\gamma_k\ket{g} = 0$, i.e. $\ket{g}$ contains no excitations:

\begin{equation}
    \ket{g} = \prod_k \left( u_k + v_kd^{\dagger}_k d^{\dagger}_{-k} \right) \ket{0},
\end{equation}
with $\ket{0}$ the vacuum of $d$ fermions. 

This procedure is great for computing the ground state, but it neglects a second degree of freedom on each $z$-bond by restricting to the sector of Hilbert space in which $\alpha_r = 1$. Dynamics will in general depend on this other degree of freedom. Therefore, in order to treat general perturbations to the magnetic Hamiltonian in Eq.\eqref{eq:1}, it is necessary to specify this further sector of Hilbert space in which $\alpha_r$ acts. To do this we note that the ground state lies in the sector for which $\alpha_r\ket{g} = \ket{g}$ for any $z$-bond. This admits of a convenient representation with a second fermion type on the $z$-bonds, analogous to the $d_r$ fermions, in terms of which $\alpha_r$ takes the form 

$$\alpha_r = 1 - 2f^{\dagger}_r f_r.$$

The relationship between $f_r$ and the Pauli matrices can be found in in Appendix \ref{app:appendixA}. In this representation, the condition that $\alpha_r = 1$ for the ground state is recast as a statement that the ground state contains no $f_r$ fermions, so that $f_r\ket{g} = 0$. The spinless fermions $f_r$ together with the $d_r$ (or the quasiparticles $\gamma_k$) give full access to the Hilbert space. It is therefore possible to write any operator acting on the lattice spin degrees of freedom in terms of $\gamma_k$, $\gamma^{\dagger}_k$, $f_r$, and $f^{\dagger}_r$. Furthermore, we have that 

$$\{ \gamma_k, f_r\} = \{ \gamma_k, f^{\dagger}_r\} = 0.$$

\section{Models for outside disturbances} \label{Models}

As discussed in Sec. \ref{Intro}, to study the coherence of the Kitaev ground state we consider four idealized models for outside disturbances to the system, representing a sudden magnetic field, a deposited impurity such as a piece of dust, global coupling to the environment (equivalent to to a noisy magnetic field), and a local dissipator (localized coupling to the environment or equivalent to a local noisy magnetic field). Measures of the coherence are obtained by treating these disturbances as quenches. 

\subsection{Magnetic field}

\begin{figure}[htp]
    \centering
    \includegraphics[width=8.5cm]{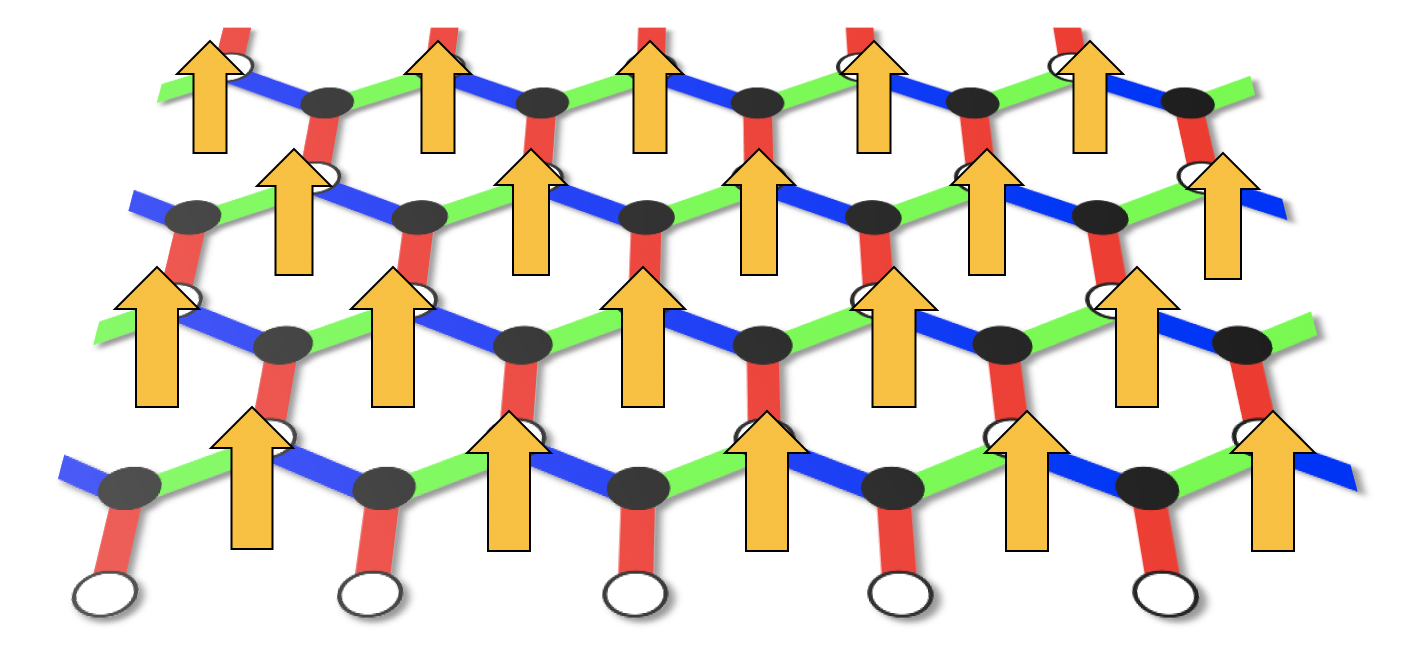}

    \caption{Kitaev honeycomb subject to a uniform magnetic field indicated by yellow arrows.  \label{fig:magfield}}
    \end{figure}
    
 Conceptually, perhaps the simplest disturbance to a magnetic system is that of a uniform magnetic field (see Fig. \ref{fig:magfield}). We treat this as a quench, imagining that a noiseless applied field is suddenly turned on at time $t = 0$ and then remains so at all future times. As in all the cases we treat here, we consider the case of a small field strength relative to the Kitaev couplings. In this regime the field can be modeled as a perturbation $V$ to the Hamiltonian so that
\begin{equation} \label{eq:6}
    H = H_0 + V,
\end{equation}
where $H_0$ is the Kitaev Hamiltonian in Eq.\eqref{eq:1} and $V$ is the potential due to a uniform magnetic field orientated in the $z$-direction, $ V = h\sum_{i} \sigma^z_{i}$.

Transforming to the $d$ and $f$ fermionic representation (see Appendix \ref{app:appendixA}) $V$ takes the form

\begin{equation} \label{eq:mag}
    V = 2h \sum_r \left(d_rf_r + f^{\dagger}_rd^{\dagger}_r \right).
\end{equation}
The term $V$ enters formally as a perturbation to the Hamiltonian - note that the disturbance is fully captured by the Hamiltonian formalism and does not incorporate noise or environmental coupling. We refer to these kinds of perturbations as being {\textit{Hamiltonian-type}}. The coherence of the ground state due to a Hamiltonian-type perturbation can be found by computing the Loschmidt echo \cite{Silva,GORIN200633,Goussev_2008}:

\begin{equation} \label{eq:echo}
     G(t) = \bra{g} e^{iH_0t}e^{-i(H_0+V)t}\ket{g}.
\end{equation}

Eq.\eqref{eq:echo} defines a measure of the separation in Hilbert space between two states: the ground state evolving via Eq.\eqref{eq:1} and an initialized ground state of $H_0$ that then evolves according to the perturbed Hamiltonian in Eq.\eqref{eq:6}.  Such a disturbance would be experimentally relevant in any situation where a weak, constant magnetic field might suddenly become coupled to the system, such as a background magnetic field from Earth.

\subsection{Impurity}

\begin{figure}[htp]
    \centering
    \includegraphics[width=8.5cm]{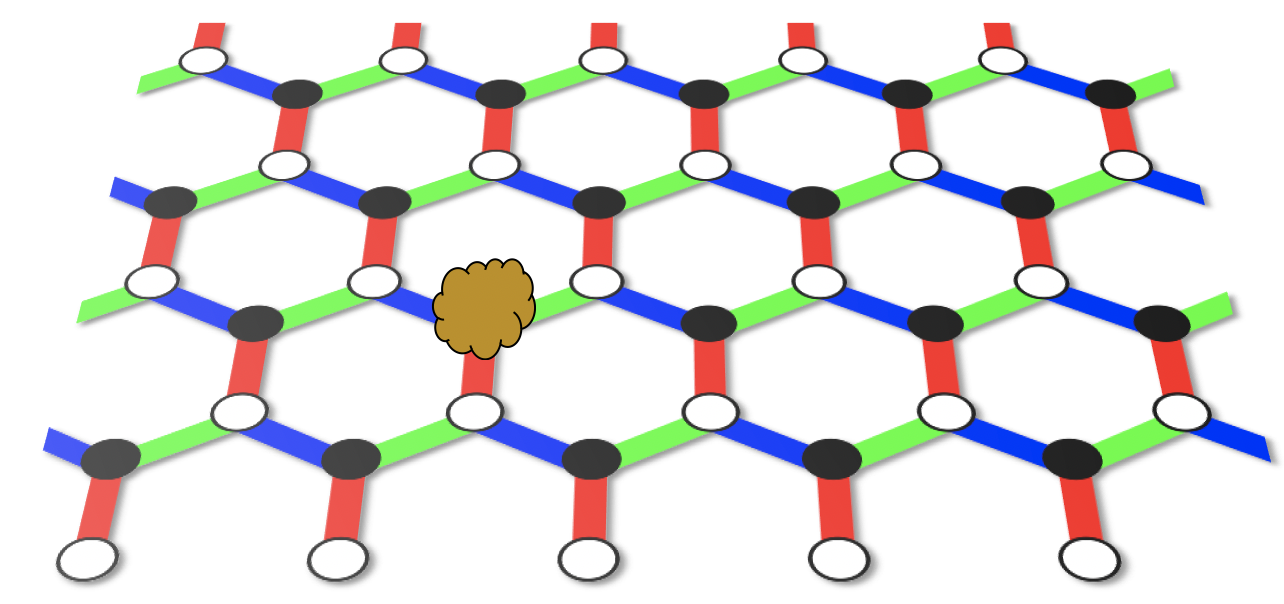}

    \caption{An impurity deposited onto the Kitaev model, shown as a brown speck of dirt, at some lattice point. We treat the case of a quenched magnetic impurity.  \label{fig:dirt}}
    \end{figure}

Another interesting case to study is that in which an impurity, perhaps a magnetic piece of dirt, is deposited somewhere in the system (see Fig. \ref{fig:dirt}). We model this situation using a noiseless magnetic impurity quench, captured in the same formalism as given above for the magnetic field, but now using a local impurity operator:
$$V = \lambda \sigma^{z}_{l},$$
where we consider an arbitrary site $l$. In the fermionic representation this can be written
\begin{equation} \label{eq:imp}
    V = \lambda \left(d_rf_r \pm d_rf^{\dagger}_r \mp d^{\dagger}_rf_r - d^{\dagger}_r f^{\dagger}_r\right).
\end{equation}
The signs of the second and third terms depend on which sublattice the impurity sits on; final results do not depend on this choice of a site. We will therefore not discuss this subtlety any further in the main text. Being that $V$ is again a Hamiltonian-type perturbation, the coherence of the ground state can be captured using the Loschmidt echo in Eq.\eqref{eq:echo}.

\subsection{Coupling to environment and noisy field}

\begin{figure}[htp]
    \centering
    \includegraphics[width=8.5cm]{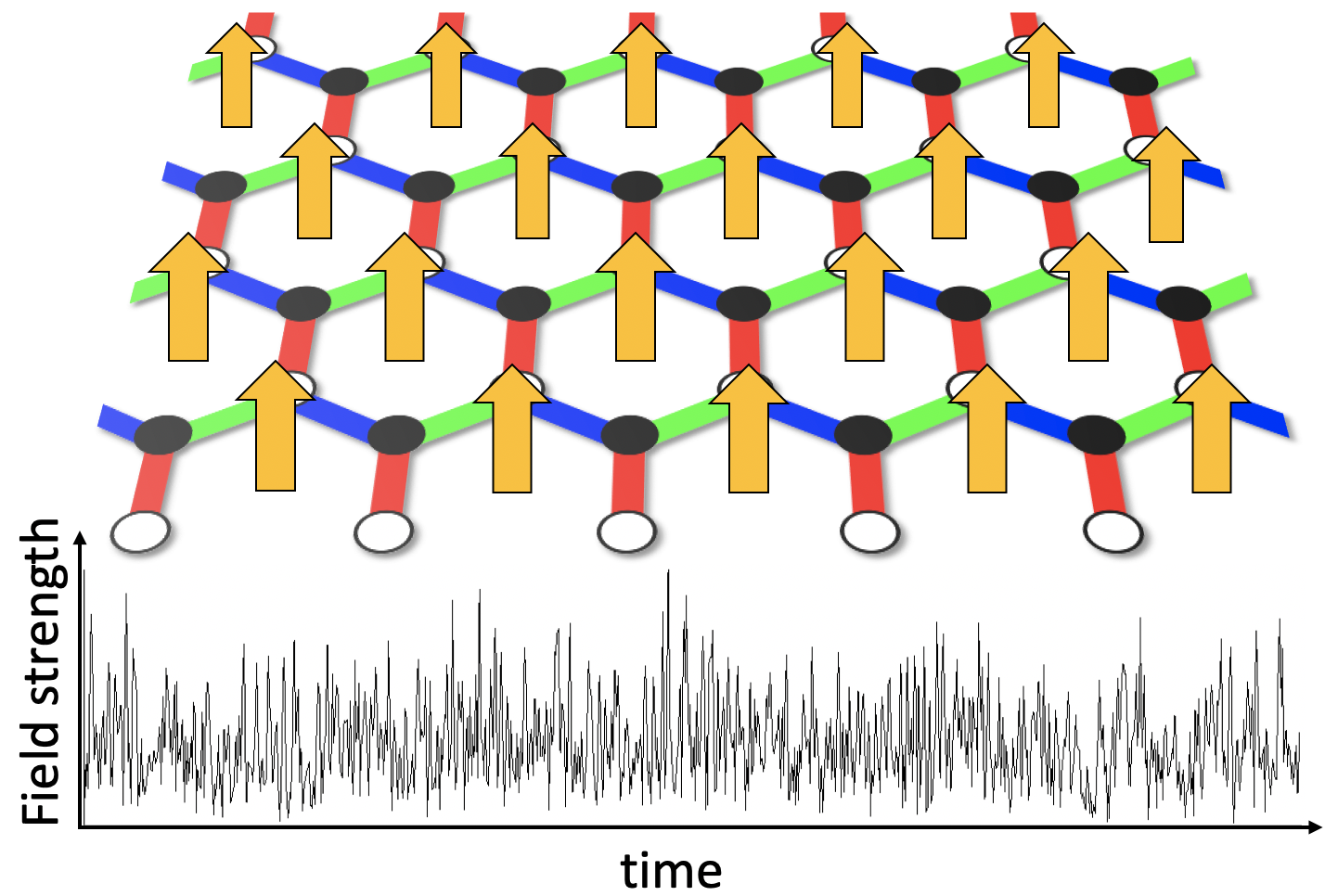}

    \caption{Kitaev honeycomb model coupled to a noisy but spatially uniform magnetic bath. We consider a uniform magnetic field subject to Gaussian white noise as a simple example of how a Lindblad jump operator can enter the quantum master equation. \label{fig:envcoupling}}
    \end{figure}
    
\begin{figure}[htp]
    \centering
    \includegraphics[width=8.5cm]{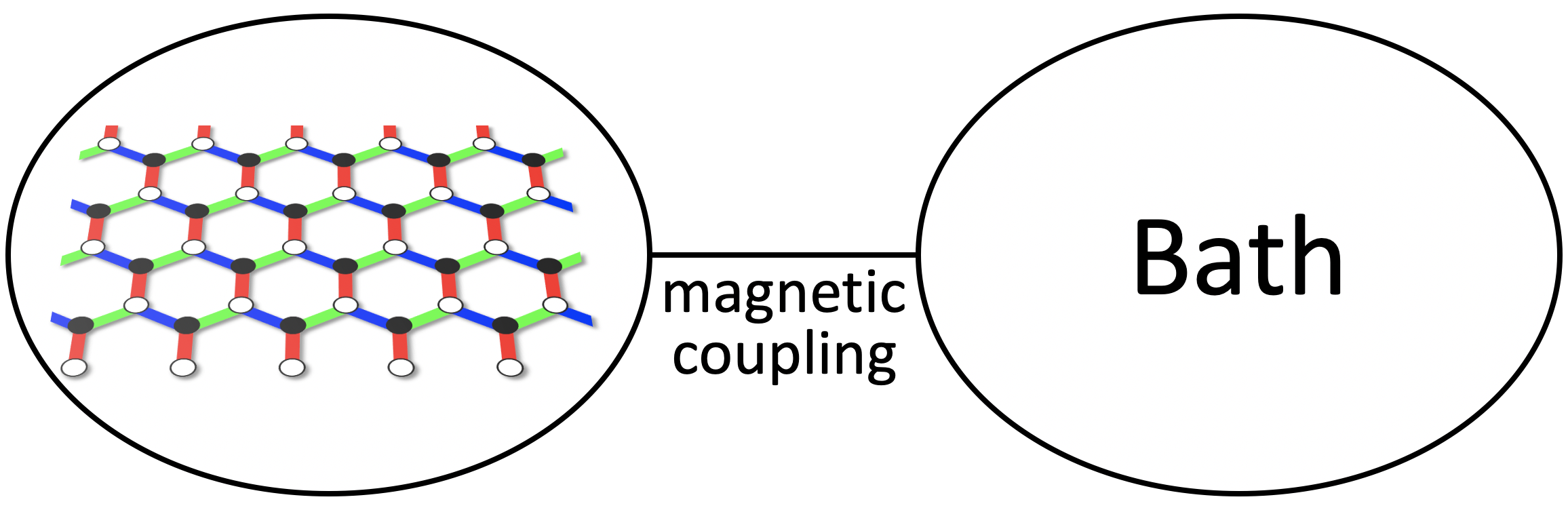}

    \caption{Coupling of the Kitaev model to a magnetic bath. Similar to the case of a noisy but spatially uniform magnetic field, coupling to a bath is a way in which the magnetic field Lindblad operator might enter the Quantum Master Equation. \label{fig:bathcoupling}}
    \end{figure}
    
Next we consider quenches that involve noise; we treat the case of Gaussian white noise. The same formalism that will be described below is also conventionally used to describe coupling to a heat bath in an open quantum system. To describe noisy disturbances to the system and bath couplings, we introduce a perturbation at the level of the Quantum Master Equation (QME) \cite{Marino}:

\begin{equation} \label{eq:QME}
    \frac{d\rho}{dt} = -i[H_0, \rho(t)] + \kappa \mathscr{L}[\rho(t)].
\end{equation}
 Here $\mathscr{L}[\rho(t)] = L\rho L^{\dagger} - \frac{1}{2} \{L^{\dagger}L, \rho\},$ and $L$ is a Lindblad operator \cite{Breuer} analogous to the $V$ discussed above, except that $L$ describes a perturbation with noise, with strength tuned by the parameter $\kappa$. 
 
 Being formulated in terms of density operators, the QME allows us to consider the general evolution of mixed states and hence incorporates statistical fluctuations and information loss, which allows for both for the treatment of noise and coupling to a heat bath. The first term on the right side of Eq.\eqref{eq:QME} is the Neumann evolution of the density matrix, and the second arises due to the Lindblad operator.
 
 The first example of such an operator that we study is a noisy uniform magnetic field (see Fig. \ref{fig:envcoupling}). Formally, there exists some sort of coupling to a magnetic bath (depicted schematically in Fig. \ref{fig:bathcoupling}) that will give rise to the same form of Lindblad operator as in the case of a noisy but spatially uniform applied field. Thus, one can interpret the results of such an analysis as pertaining to both situations.  As magnetic shielding becomes more technologically reliable, weak coupling will be the relevant regime to consider. Hence, to capture environmental coupling or a weak noisy field, we study the Lindblad operator \cite{PhysRevLett.118.140403} 
\begin{equation}
    L = h \sum_{i} \sigma^z_{i} = 2h \sum_r \left(d_rf_r + f^{\dagger}_rd^{\dagger}_r \right) \label{eq:noisyfieldop}
\end{equation}
just as in the case of the noiseless applied field. Considering the weak-coupling case amounts to taking $\kappa h^2$ to be small relative to the Kitaev couplings.
While the Loschmidt echo in Eq.\eqref{eq:echo} works for studying the evolution of pure states, to study noise we need a generalized coherence measure that can capture evolution to mixed states, and hence is written in terms of density operators. If we again consider quenches, in which case the noisy disturbance is suddenly turned on, the coherence of a quantum state evolving under Eq.\eqref{eq:QME} may be captured via the Uhlmann fidelity \cite{Marino}:

\begin{equation} \label{eq:fidtrace}
    F(t) = \Tr[\rho_0(t) \rho(t)],
\end{equation}
where the density matrices are defined so that  $\rho_0(t)$ evolves under $L = 0$ and $\rho(t)$ evolves according to the full QME in Eq.\eqref{eq:QME}, and $\rho_0(0) = \rho(0) = \ket{g}\bra{g}$. The structure is completely analogous to the Loschmidt echo, giving the overlap of two density operators, one being the pure ground state density operator evolved according to the Kitaev Hamiltonian, and the other being the ground state initialized at time $t = 0$ before being evolved under the presence of the noisy disturbance. In fact for a Hamiltonian-type disturbance, the Uhlmann fidelity and Loschmidt echo are related by $F(t) = |G(t)|^2$

\subsection{{Local dissipator}}

\begin{figure}[htp]
    \centering
    \includegraphics[width=8.5cm]{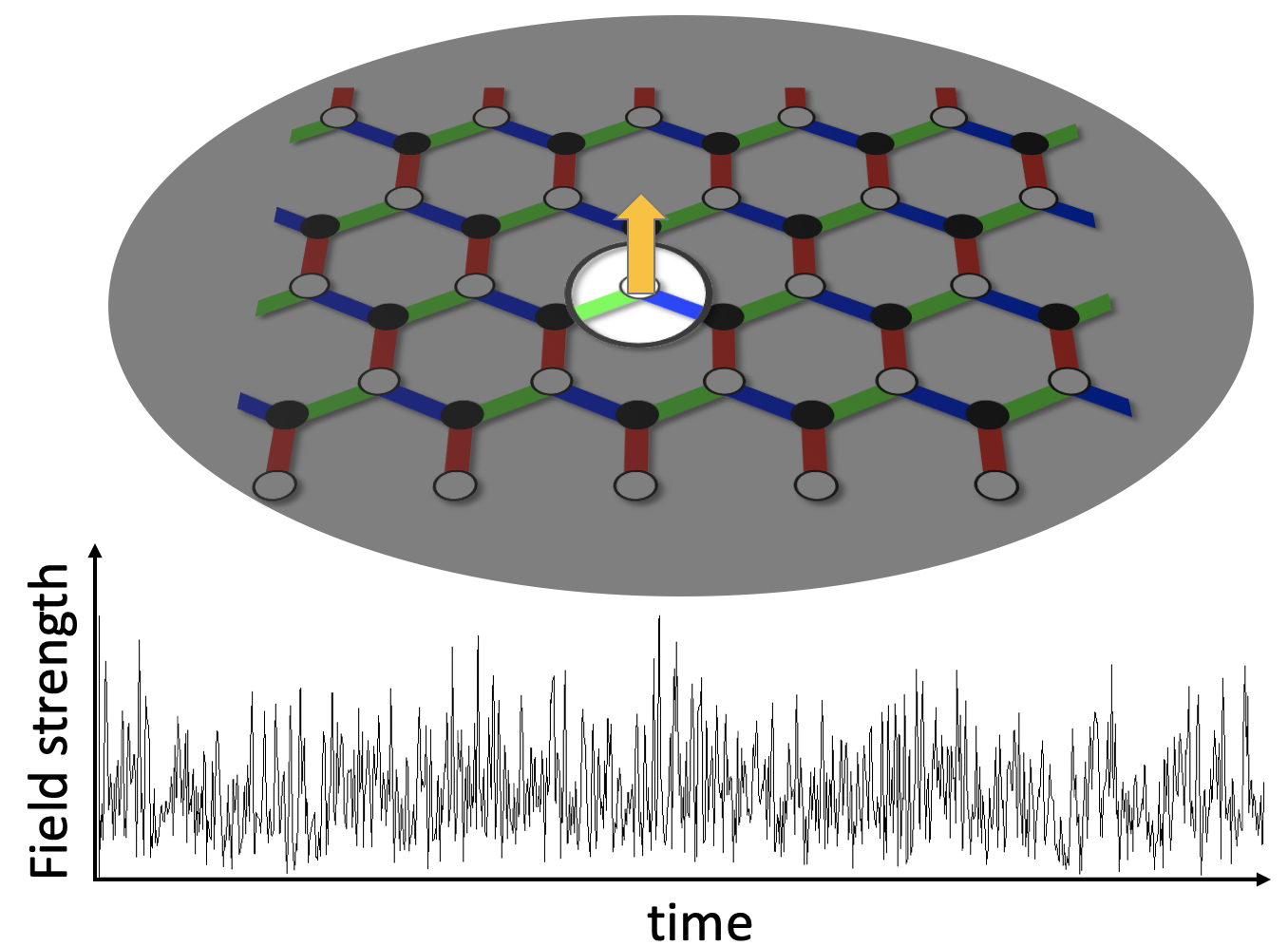}

    \caption{A hole in the system's magnetic shielding couples it locally to a noisy magnetic environment. Information can then leak from the system and be lost to the environment. We model this situation using a noisy magnetic impurity quench.  \label{fig:dissipator}}
    \end{figure}

This formalism may also be used to model a scenario in which information is lost from the system locally. For example, we can imagine a case in which a hole appears in the shielding of a device, coupling it locally to the environment or allowing magnetic noise to enter (see Fig. \ref{fig:dissipator}). We treat this situation using a local dissipator via a Lindblad operator as in Eq.\eqref{eq:QME} \cite{PhysRevLett.118.140403}. This is modeled using a magnetic impurity with Gaussian white noise, so that the Lindblad operator is
\begin{equation}
    L =  \lambda \sigma^{z}_{l} = \lambda \left(d_rf_r \pm d_rf^{\dagger}_r \mp d^{\dagger}_rf_r - d^{\dagger}_r f^{\dagger}_r\right).
    \label{eq:noisy_imp_L}
\end{equation}
As in the case of the noiseless impurity in Eq.\eqref{eq:imp}, the middle terms have signs that depend on the sublattice where the dissipator exists, but we find this sign does not have any physical consequences. We will therefore not discuss this subtlety further, as in the noiseless case. 
To capture the robustness of the Kitaev ground state to the local dissipator, the fidelity in Eq.\eqref{eq:fidtrace} is computed.

\section{Mathematical approach} \label{sec:math}

All cases outlined in Sec. \ref{Models} will be treated using a cumulant expansion of the Loschmidt echo and the Uhlmann fidelity to second order \cite{Silva,Marino}. Details of the cumulant expansion are given in Appendix \ref{ApExp}. Here we simply remark that the cumulant expansion is a partial resummation of a more straightforward perturbative expansion, and is constructed to capture exponential behavior particularly well. 

\subsection{Cumulant expansion for Loschmidt echo}

For the noiseless quenches we study, we wish to compute the Loschmidt echo, which is an expectation value of an operator exponential, introduced in Eq.\eqref{eq:echo}. We note at the outset that matrix exponentials of operators involving terms that are quartic in fermionic operators are in general difficult to compute. 
With this in mind, we recall that in the $d$ and $f$ representation introduced above, the Kitaev Hamiltonian in Eq.\eqref{eq:1} takes the form

\begin{dmath}
     H_0 = \sum_r J_x(d^{\dagger}_r + d_r)(d^{\dagger}_{r+ex} - d_{r+ex}) + \sum_r J_y(d^{\dagger}_r + d_r)(d^{\dagger}_{r+ey} - d_{r+ey}) + \sum_r J_z(1 - 2f^{\dagger}_rf_r)(2d^{\dagger}_r d_r - 1),
\end{dmath}
which indeed contains a quartic  $f^{\dagger}_rf_rd^{\dagger}_r d_r$ term. As mentioned above these kinds of terms make it difficult to compute operator exponentials. However, as we will see below it is necessary to compute matrix exponentials of a non-perturbed Hamiltonian to be able to  perturbatively compute the Loschmidt echo in Eq.\eqref{eq:echo} for a perturbed Hamiltonian of the form $H = H_0 + V$, where $V$ will be given by Eq.\eqref{eq:mag} or Eq.\eqref{eq:imp}. We wish to avoid complications in the calculation arising from the quartic term $\propto f^{\dagger}_rf_r(2d^{\dagger}_rd_r - 1)$. To achieve this it is expedient to recognize that $H_0\ket{g} = H_1\ket{g}$ for an auxiliary Hamiltonian 

\begin{dmath}
H_1 = \sum_r J_x(d^{\dagger}_r + d_r)(d^{\dagger}_{r+ex} - d_{r+ex}) + \sum_r J_y(d^{\dagger}_r + d_r)(d^{\dagger}_{r+ey} - d_{r+ey}) + \sum_r J_z(2d^{\dagger}_r d_r - 1).
\end{dmath}
This is the system that was solved by setting $\alpha_r = 1$ in Eq.\eqref{Halpha}. The choice of $H_1$ as non-perturbed Hamiltonian rather than $H_0$ has the advantage that it involves only bilinear terms. Therefore, it is convenient to treat the problem in terms of this Hamiltonian and push the quartic operator into the perturbation, so that

\begin{equation} \label{eq:vtilde}
\tilde{V} =V -2J_z \sum_r f^{\dagger}_rf_r(2d^{\dagger}_rd_r - 1).
\end{equation}

Naively, for any perturbative analysis this approximation should lock us into a corner of the Kitaev phase diagram. That is, because $\tilde{V}$ must be small relative to $H_0$ for a perturbative treatment to be reliable, it stands to reason that in so defining $\tilde{V}$ as the perturbation one should only consider regions in parameter space where $J_z \ll J_x, J_y$. However, the approximation could be valid for larger $J_z$ due to the presence of $f^{\dagger}_rf_r$, especially when one is studying properties of the ground state for which $f^{\dagger}_rf
_r\ket{g} = 0$. If the expectation value of $f^{\dagger}_rf_r$ is small for the states relevant to the calculation, then $\tilde{V}$ could be small even for $J_z \sim \mathcal{O}(J_x)$. As it turns out, this is so for our calculations; to second order in the cumulant expansion outlined below and detailed in Appendix \ref{ApExp}, the term $\propto f^{\dagger}_r f_r$ in $\tilde{V}$ makes no contribution to the expectation values, and the approximation is therefore valid even for large $J_z$.

With this repartitioning of terms, it makes sense to write the Loschmidt echo as 
\begin{equation} 
\begin{aligned}
G(t) &=\bra{g}e^{iH_0t}e^{-i(H_0+V)t}\ket{g}\\
    &= \bra{g}e^{iH_1t}e^{-i(H_1+\tilde{V})t}\ket{g}.
    \end{aligned}
    \label{eq:GH1}
\end{equation}
where in the second equality we have used the fact that $H_1|g
\rangle=H_0|g\rangle$. Working in the interaction picture with $H_1$ taken as the non-interacting Hamiltonian allows us to rewrite the expression as 
\begin{equation}
    G(t) = \bra{g} Te^{-i\int^t_0 dt' \tilde{V}(t')}\ket{g},
\end{equation}
where $\tilde{V}(t) = e^{iH_1t}\tilde{V}e^{-iH_1t}.$

In our case a naive perturbative expansion for small $\tilde{V}(t)$ is best resummed into a cumulant expansion \cite{Marino,PhysRevB.94.014310} because it allows for a more convenient way to capture exponential behaviour - which is useful in the description of decay at relatively long time scales. To second order we find that

\begin{equation} \label{echoapp1}
    G(t) \approx e^{-i\int^t_0 dt' \left<\tilde{V}(t')\right>^c_0}e^{-\frac{1}{2}\int^t_0 \int^t_0 dt_1 dt_2 \left<T\tilde{V}(t_1)\tilde{V}(t_2)\right>^c_0}.
\end{equation}

As mentioned above, we find that regardless of the size of $J_z$ the operator $-2J_z \sum_r f^{\dagger}_rf_r(2d^{\dagger}_rd_r - 1)$ does not modify either cumulant. That is, $\left<\tilde{V}(t')\right> = \left<V(t')\right>$ and $\left<T\tilde{V}(t_1)\tilde{V}(t_2)\right>^c_0 = \left<TV(t_1)V(t_2)\right>^c_0$, where $V(t) = e^{iH_1t}Ve^{-iH_1t}$ is the quench operator in the interaction picture. This suggests that to second order in the cumulant expansion it is consistent to shift the $f$-dependent operator into the perturbation even for large $J_z$.

Using this expansion, we find the Loschmidt echo in Eq.\eqref{echoapp1} for the impurity case,  Eq.\eqref{eq:imp}. We refer the interested reader to Appendix \ref{B} for details about the calculations of the first and second cumulants. The result is  

\begin{dmath} \label{fullimpecho}
     |G_{l}(t)| \approx \exp{\left[\lambda^2\int_{-\pi}^{\pi}\frac{d^2k}{(2\pi)^2} \frac{\cos{(E_kt)}-1}{E^2_k}\right]},
\end{dmath}
where we consider only the absolute value because this is the piece that determines the nontrivial time dependence of the Loschmidt echo.

For the case of the magnetic field we find from Eq.\eqref{eq:mag} and Eq.\eqref{echoapp1} that 
\begin{dmath} \label{fullmagecho}
     |G_{u}(t)| \approx  \exp{\left[4Ah^2\int_{-\pi}^{\pi}\frac{d^2k}{(2\pi)^2} \left[ \cos{(E_kt)} -1\right]\frac{|u_k|^2}{E^2_k}\right]},
     \label{eq:mag_field_loschmidt}
\end{dmath}
where $A$ is the system size. We note that even before computing the momentum integrals, one can see that for the magnetic field case there will be a dependence on the system size, while for the impurity quench this is not so.

For general choices of parameters $J_x$, $J_y$, and $J_z$ integrals appearing in both expressions are not easily computed. This is especially obvious for the gapless phase where poles appear at $E_k = 0.$ To gain further analytical insight into our results we need to restrict ourselves to a physically interesting regime. The least well studied regime that is often expected to play host to interesting universal phenomena is the long-time regime \cite{Marino}. This regime will also become more experimentally relevant as coherence times grow in the field of quantum information. Therefore, we will approximate Eq.\eqref{fullmagecho} and Eq.\eqref{fullimpecho} assuming a long-time regime. This is achieved by applying a stationary phase approximation in the gapped case and a more sophisticated approach that involves fitting in the gapless case. Details of these methods are discussed in Appendix \ref{app:C}.

\subsection{Cumulant expansion for Uhlmann fidelity}

As discussed above, for cases involving noise we study coherence through the Uhlmann fidelity \eqref{eq:fidtrace}, which enables us to treat a noisy quench by analyzing the evolution of density operators rather than kets. One may express the fidelity using a so-called superoperator formalism \cite{Marino}. This choice of formalism has the advantage that a second-order cumulant expansion proceeds analogously to that for the Loschmidt echo (see Appendix \ref{ApExp}). More precisely, in the superoperator formalism density operators are treated as generalized vectors, and mappings from operators to operators are written as generalized operators \cite{Marino}. Concretely, this means that we will use the dictionary:
$$\rho \rightarrow |\rho) $$
$$\mathscr{L}[\rho] = \mathscr{L}|\rho)$$
$$\Tr[A^{\dagger}B] = (A|B)$$
for linear operation $\mathscr{L}$ taking matrices to matrices, and inner products of square matrices $A$ and $B$. In this formalism the fidelity can be recast as an inner product
\begin{equation} \label{fidelity}
    F(t) = \Tr[\rho_0(t) \rho(t)] = (\rho_0(t)|\rho(t)).
\end{equation}
The structure of Eq.\eqref{fidelity} is very similar to the Loschmidt echo: an inner product between a state evolved via unperturbed dynamics and one with perturbation. The key difference is that here the evolution happens according to a Lindblad QME in Eq.\eqref{eq:QME}.
As before, it is convenient to separate the $f$-dependent piece from the Hamiltonian so that $H_0 = H_1 + W$ and $W = -2J_z \sum_r f^{\dagger}_rf_r(2d^{\dagger}_rd_r - 1).$

$V$ may be viewed as a Hamiltonian-type perturbation, without noise. The QME becomes
\begin{dmath}
    \frac{d\rho}{dt} = -i[H_1, \rho(t)] -i[W, \rho(t)] + \kappa \mathscr{L}[\rho(t)] \equiv  -i[H_1, \rho(t)] +  \tilde{\mathscr{L}}[\rho(t)].
\end{dmath}

Working in the interaction picture where $L_I(t) = e^{iH_1 t} L e^{-iH_1 t}$ and $W_I(t) = e^{iH_1 t} W e^{-iH_1 t}$, we define $\tilde{\mathscr{L}}_{I}(t)[\rho]$ to be the super-operator generated by $W_I$ and $L_I$. From here we perform a second-order cumulant expansion on the fidelity written in the interaction picture, 

$$F(t) = \Tr[\rho_0(t) \rho{t}] = (\rho_0 | \rho) = (\rho_0(0)|\rho_I(t)),$$ where $\rho_0$ is time evolved according to $H_1$ and $\rho$ is evolved according to the full Master Equation. We are interested in the stability of a pure ground state $\rho_0(0) = \rho(0) = \ket{g}\bra{g}$. Here, the Lindblad operator after a finite time makes it possible for the system to evolve from the pure state to a mixed state.

In the superoperator formalism, the second-order cumulant expansion proceeds in direct analogy to that for the Loschmidt echo, giving
\begin{equation}
    F(t) = \exp{\left[\int^t_0 ds (\tilde{\mathscr{L}}_{I}(s))_0 + \frac{1}{2} \int^t_0 \int^t_0 ds ds' (T \tilde{\mathscr{L}}_{I}(s) \tilde{\mathscr{L}}_{I}(s'))^c_0\right]},
\end{equation}
with $(\tilde{\mathscr{L}}_{I}(s))_0 = (\rho_0|\tilde{\mathscr{L}}_I(s)|\rho_0)$ and $(T \tilde{\mathscr{L}}_{I}(s) \tilde{\mathscr{L}}_{I}(s'))^c_0 = (T \tilde{\mathscr{L}}_{I}(s) \tilde{\mathscr{L}}_{I}(s'))_0 - (\tilde{\mathscr{L}}_{I}(s))_0(\tilde{\mathscr{L}}_{I}(s'))_0$. As before, we find that to second order the $f$-dependent term $W$ does not contribute to the expansion for either the noisy field or noisy impurity. This also {\it a posteriori} justifies our perturbative treatment of these terms as small - although it is not immediately obvious that they would be small by naive considerations. 
%It is therefore equivalent up to this order to compute using $\mathscr{L}$ instead of $\tilde{\mathscr{L}}$, again reinforcing the treatment of this term that is not necessarily small as part of the perturbation. 

Details of the calculation of these cumulants for the systems studied can be found in Appendix \ref{B}. The fidelity for the local dissipator in Eq.\eqref{eq:noisy_imp_L} becomes

\begin{equation} \label{Eq:localfid}
    F_l(t) = e^{-\kappa t } \exp{\left[\frac{\kappa^2}{2} \int \frac{d^2k}{(2\pi)^2} \frac{d^2k'}{(2\pi)^2} \frac{1 - \cos{((E_k - E_{k'})t)}}{(E_k - E_{k'})^2}\right]}.
\end{equation}
%\mvcomment{Before going to enirvonmental couplings I would suggest maybe discussing something you can already see in the result like the dependence on $\kappa$ and no area term entering. Alternatively, you could contrast this to the "coupling to environment result" after showing that result.}
Note that even without computing the integral we see that the leading behavior (from the $\mathcal{O}(\kappa)$ term) is exponential decay tuned by the coupling $\kappa$, and there is no dependence on system size. 
For the environmental coupling in Eq.\eqref{eq:noisyfieldop} we find
\begin{equation} 
    F_u(t) = \exp{\left[-4\kappa h^2 A t \int \frac{d^2k}{(2\pi)^2} |u_k|^2\right]} \exp{\left[\alpha(t)\right]},
\end{equation}
with $ \alpha(t) =16\kappa^2 h^4 \times $
\begin{dmath} \label{Eq:ufid}
     \left[ \frac{1}{2}t^2 A\int{ \frac{d^2k}{(2\pi)^2} |v_k|^2|u_k|^2}  - A^2\int{\int \frac{d^2k}{(2\pi)^2} \frac{d^2k'}{(2\pi)^2} |u_k|^2|u_{k'}|^2 \frac{1 - \cos{(E_k + E_{k'})t}}{(E_k + E_{k'})^2}} + A\int{ \frac{d^2k}{(2\pi)^2}|u_k|^2\frac{1 - \cos{2E_kt}}{(2E_k)^2}} + A^2 \int{\int \frac{d^2k}{(2\pi)^2} \frac{d^2k'}{(2\pi)^2} |u_k|^2|u_{k'}|^2 \frac{1 - \cos{(E_k - E_{k'})t}}{(E_k - E_{k'})^2}} \right],
\end{dmath}
where A is the system size. As in the local case the first cumulant gives exponential decay.  However in the case of the noisy field the strength of this decay depends on system size, noise strength $\kappa$, field strength $h$, and the Kitaev couplings through $|u_k|^2$. The presence of a positive $t^2$ term in the second cumulant signals a breakdown of the cumulant expansion on some timescale depending on the sizes of $\kappa$, $h$, and $A$. However, for a large system size ($A\to \infty$), it is reasonable to drop the terms linear in $A$ and keep only those $\propto A^2$ in the second cumulant. This gives the simplified expression
\begin{dmath} \label{Eq:alpha}
      \alpha(t) =16\kappa^2 h^4 A^2 \times \left[-\int{\int \frac{d^2k}{(2\pi)^2} \frac{d^2k'}{(2\pi)^2} |u_k|^2|u_{k'}|^2 \frac{1 - \cos{(E_k + E_{k'})t}}{(E_k + E_{k'})^2}} + \int{\int \frac{d^2k}{(2\pi)^2} \frac{d^2k'}{(2\pi)^2} |u_k|^2|u_{k'}|^2 \frac{1 - \cos{(E_k - E_{k'})t}}{(E_k - E_{k'})^2}}\right],
\end{dmath}
which is reasonable on timescales for which the quadratic term is small relative to the first cumulant, i.e. on the order of $t \lesssim \frac{1}{2\kappa h^2}$. Moreover, restricting $\kappa h^2$ to be sufficiently small relative to the energy scale of the unperturbed system ensures that even for large $A$, the first cumulant will be larger than the second.

The leading-order behavior in both cases is exponential decay, a manifestation of Anderson's Orthogonality Catastrophe \cite{AndersonOrtho}. While in the impurity case this decay depends only on the noise strength $\kappa$, for the noisy field we have leading order $\exp{(-{\rm const.}\times A \kappa h^2 t)}$. This immediately suggests more rapid decay for larger systems when environmental coupling is present. 

\subsection{Integral approximations}
While we have computed cumulant expansions in the previous sections, these results still include momentum integrals that have to be computed. The purpose of this section is to summarize the methods we will employ to compute those integrals.

We will first focus on the simpler case of the gapped phase, where $E_k \neq 0$ for all $k$. This structure permits us to simplify integrals with a factor $1/E^2_k$ or $1/(E_k + E_{k'})^2$ in the integrand. Integrals can be computed via a stationary phase approximation for long times, as discussed in Appendix \ref{app:C}. That is, it is a good approximation to apply the general formula 
\begin{dmath} \label{spapp} \label{spformula}
    \int^{\pi}_{-\pi} \frac{d^2k}{(2\pi)^2} f(\vec{k})\cos{(\phi(\vec{k}) t)} \approx \operatorname{Re} \sum_{k_0} \left(\frac{1}{2}\right)^p (-1)^{\delta_{p-2}}e^{i\phi(\vec{k}_0)t}f(\vec{k}_0)\frac{i}{2\pi t}(\operatorname{det}\phi'')^{-1/2},
\end{dmath}
which gives the long-time behavior to leading order. The sum is over stationary points $\vec{k}_0$ of the function $\phi(\vec{k})$. The variable $p$ is given as $p=0$ for $\vec{k}_0 = (0,0)$, $p=1$ for stationary points on the edge of the BZ such as $\vec{k}_0 = (0,\pi)$, and $p=2$ for stationary points at corners such as $\vec{k}_0 = (\pi, \pi)$. 

In the gapless case the stationary phase approximation does not apply directly because the asymptotic behavior of the integrals depends not only on stationary phase regions of $k$-space, but also on regions near poles where the denominator of the integrand is zero. In such cases we find that the stationary phase formula in Eq.\eqref{spformula} is still useful if it is applied to second or third derivatives in time, which are themselves free of such poles and for which the stationary phase approximation is good. To approximate $I(t)$ appearing in $|G(t)| = \exp{(I(t))}$ or $F(t) = \exp{(I(t))}$ in this case, we carry out this approximation for the higher derivative $\partial^n_t I$, then integrate the result $n$ times to produce a result for $I(t)$, fitting the integration constant numerically at each subsequent integration. More details on this process can be found in Appendix \ref{app:C}. 

\section{Results} \label{sec:results}

In this section we discuss explicit expressions of the Loschmidt echo and fidelity for the different types of quenches we consider. We expect that gapped and gapless systems will display fundamentally different behavior. However, there are not many fundamental changes between different gapped cases with different couplings $J_i$, nor are there many differences between the different gapless cases. Therefore, to reduce the complexity of analytical expressions in the results, we choose one representative point for coupling parameters $J_i$ in the gapless phase and one in the gapped phase. 

Therefore, for each type of quench we study two characteristic choices of the coupling parameters $J$, with $\vect J=(J_x,J_y,J_z)=(1,1,1)$ giving a gapless dispersion and $\vect J=\sqrt{3/11}(1,1,3)$ giving a gapped dispersion. The parameters for the gapped system are chosen so that the Hamiltonians for the gapped and gapless system have the same Frobenius norm. This ensures that the Loschmidt echo or fidelity for gapped and gapless cases can be compared more easily. That is, it ensures that differences in $G(t)$ or $F(t)$ between gapped and gapless cases are not due to differences in energy scale. Results for the long-time behavior of the coherence are summarized in  Table \ref{Tab:res1}. In particular we highlight the structural similarities of the coherence measures across cases, as well as the specific dependence on parameters such as field strength and system size.

\begin{table*}[t]
  
\begin{center}
\centering
\begin{tabular}{| >{\centering}p{2.5cm} || >{\centering}p{1cm} | >{\centering}p{1cm} || >{\centering}p{1cm} | >{\centering}p{1cm} || >{\centering}p{2.5cm} | >{\centering}p{1cm} || >{\centering}p{3.7cm} | >{\centering}p{1.5cm} ||} 
\hline 
$ F(t) \sim Ce^{-\alpha t} t^{-\beta}$&  \multicolumn{2}{c||}{impurity}  &  \multicolumn{2}{c||}{magnetic field} &  \multicolumn{2}{c||}{noisy impurity}  & \multicolumn{2}{c||}{environmental coupling} \tabularnewline [1ex]
 \hline
 & $\alpha$ &  $\beta$ &  $\alpha$ &  $\beta$ &  $\alpha$ & $\beta$ & $\alpha$ & $\beta$ \tabularnewline[1ex]
 \hline
  Gapped & $0 $ & $ 0 $  & $ 0 $& $ 0 $ & $ \kappa - \kappa^2 \times  {\rm const.} $& $\propto \kappa^2 $ & $\propto A h^2 \kappa - A^2 h^4 \kappa^2 \times  {\rm const.} $ & $\propto A^2 h^4 \kappa^2$ \tabularnewline[1ex]
 \hline
 Gapless & $ 0 $ & $\propto \lambda^2 $ & $ 0$ & $\propto Ah^2 $ & $ \kappa - \kappa^2 \times  {\rm const.} $ & $\propto \kappa^2$ & $\propto A h^2 \kappa - A^2 h^4 \kappa^2 \times  {\rm const.}$& $\propto A^2 h^4 \kappa^2 $\tabularnewline[1ex] 
 \hline
\end{tabular}
\end{center}

\begin{tabular}{|c|c|}
   \hline  $\lambda$ & impurity strength \\
  \hline $A$  & system size \\
  \hline $h$ & magnetic field strength \\
  \hline $\kappa$ & noise strength \\
  \hline
\end{tabular}

\caption{Asymptotic behavior ($t\to\infty$) of the Uhlmann fidelity, $F(t)$, measuring the coherence of the Kitaev ground state when a quench is applied. For noiseless cases we can identify $F(t) = |G(t)|^2$. Quenches studied are a magnetic impurity, magnetic field, noisy magnetic impurity, and environmental coupling, each for both gapped and gapless models. For the noisy cases a universal form $Ce^{-\alpha t} t^{-\beta}$ is found, where the specific form of $\alpha$ and $\beta$ depends on the type of quench. $C$ is constant in the long-time limit. The magnetic impurity is modeled by a perturbation $V = \lambda \sigma^z_l$ and the magnetic field by $V = h \sum_i \sigma^z_i$, where the sum is over sites in the honeycomb lattice. Noise is a treated using a generalization of these operators in a Lindblad formalism, where a small parameter $\kappa$ determines the strength of the noise. \label{Tab:res1}}
\end{table*}

\subsection{Loschmidt echo under the magnetic field quench}
We begin by discussing the effect that a quenched uniform magnetic field will have on the ground state of the Kitaev honeycomb model.
\subsubsection{Gapped system}
We first consider the gapped system with $\vect J= \sqrt{3/11}(1,1,3)$. For the gapped system Eq. \eqref{spformula} can be applied directly to Eq. \eqref{fullmagecho} for the Loschmidt echo, leading to 
\begin{dmath}
    |G_u(t)| \approx e^{-c_u A h^2} \exp{\left[ \frac{4Ah^2}{\pi t} \left(-\frac{11^{3/2}}{144}\sin{(2\sqrt{\frac{3}{11}}t)}\right)\right]}\times \exp{\left[\frac{4Ah^2}{\pi t} \left( \sqrt{\frac{11}{3}}\frac{11}{216} \cos{(6\sqrt{\frac{3}{11}}t)}\right) \right]} \times \exp{\left[\frac{4Ah^2}{\pi t} \left( \frac{11\sqrt{55}}{3600}\sin{(10\sqrt{\frac{3}{11}}t)}\right)\right]},
\end{dmath}
which is valid at intermediate to long timescales. As $t\to\infty$ we see that $|G_u| \to e^{-c_u A h^2}$ where $c$ is a constant

$$c_u = 4\int_{-\pi}^{\pi}\frac{d^2k}{(2\pi)^2} \frac{1}{E^2_k}.$$

For the case of $\vect J=\sqrt{3/11}(1,1,3)$ we have $c_u = 0.5278$. We therefore find that for the gapped system under an impurity quench, the size of $|G_u(\infty)| = e^{-0.5278 A h^2}$ is determined fully by the field strength $h$ and the system size $A$. That is, it is time-independent and corresponds to a finite overlap with the Kitaev ground state as time $t \to \infty$. Notably, an Orthogonality Catastrophe does not manifest in this case. In particular, the size of this finite coherence at long times may be tuned via sample area size $A$ and coupling strength $h$, where larger $Ah^2$ leads to reduced coherence. While to ensure consistency with a cumulant expansion we assume that $h$ is small relative to the Kitaev couplings, there is no such restriction on system size. Thus, we find that for fixed $h$, larger system size will correspond to reduced coherence, while smaller system size will enhance coherence.  
%\mvcomment{Maybe say something about that ystem sizes $A$ and coupling $\lambda$ can in this case be directly determined if one wants to engineer a system that that reaches a minimum specific value of Loschmidt echo. If you can find a direct application of this it can make things sound very interesting.}

We should also stress that the next-to-leading-order $\exp(1/t)$ decay profile for a gapped system under both impurity and magnetic field quenches is a universal feature of the gapped phase. This is because it is determined solely by the number of Brillouin zone integrations and the dimensionality of $k$-space. For an n-dimensional $k$-space, application of a stationary phase approximation to an integrand such as that in Eq.\eqref{spformula} will produce a power of $(t)^{-n/2}$ for the Loschmidt echo \cite{Bender}. Afterall, each 1D Gaussian integral in the $k$-space integrals of the stationary phase approximation produces a factor $t^{-1/2}$. The asymptotic decay to a constant  under a noiseless quench can therefore be expected to be a universal feature of the gapped phase, which does not crucially depend on a specific dispersion.

\subsubsection{Gapless system}
To contrast the behavior of the gapped phase we next consider the gapless case with $\vect J = (1,1,1)$. Because in the gapless phase $E_k = 0$ for some points in the BZ, the stationary phase approximation cannot be directly applied to the integrals in any of the expressions for $G(t)$ or $F(t)$,  all of which have integrands containing $E_k$ to some power in the demoninator. More precisely, contributions near stationary phase points will only contribute sub-dominant long-time behavior. The dominant behavior at long times arises due to contributions near singular points of $1/E_k$. It should be stressed that contributions from the singular points of $1/E_k$ to the full integrals in Eq.\eqref{fullmagecho} will be finite for finite $t$, as it can seen by a Taylor expansion of the full integrand 
$$\frac{\cos{(E_kt)}-1}{E^2_k} = -\frac{1}{2} t^2 + \mathcal{O}(E^2_k).$$

Details about the methodology for finding the asymptotic behavior of these integrals in the gapless phase is discussed in Appendix \ref{app:C}. Here, we merely provide a brief summary.
Considering $|G(t)| = \exp{(4Ah^2I(t))}$ for 
$$I(t) = \int_{-\pi}^{\pi}\frac{d^2k}{(2\pi)^2} \left[ \cos{(E_kt)} -1\right]\frac{|u_k|^2}{E^2_k},$$
we leverage the fact that a stationary phase approximation will capture the long-time behavior of some higher time derivative $\frac{d^n}{dt^n}I$. In this case applying two time derivatives removes the $E^2_k$ from the integrand, and we can apply a straightforward stationary phase approximation to $\frac{d^2}{dt^2}I$ via Eq.\eqref{spformula}. We then employ additional semi-analytical steps to extract the behavior of $I(t)$ from $\frac{d^2}{dt^2}I$, as detailed in Appendix \ref{app:C}.

%\mvcomment{The last two sentences need to be clearer. It is a strange combination of very specific and vague at the same time. One thing you could do is just be even vaguer by just stating that additional semi-analytic steps are performed to extract the behaviour of I(t) from its derivatives, which are detailed in the appendix.} 

For the uniform field this process gives
\begin{dmath}
    |G_u(t)| \approx \exp{\left[A h^2(-c_{h0} -c_{h1} \log{t} + \mathcal{O}(1/t))\right]}, 
\end{dmath}
where a fit gives the numerical values of $c_{h0} = 0.586$, and $c_{h1} = 0.183 $. The $\mathcal{O}(1/t)$ term is the contribution to $I(t)$ from regions of stationary phase. 

\begin{figure}[htp]
    \centering
    \includegraphics[width=8.5cm]{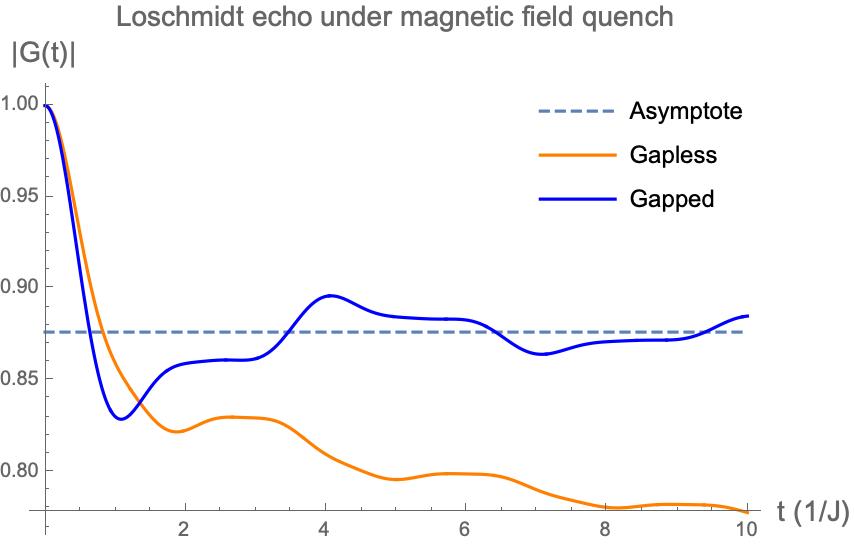}

    \caption{Numerical plots of the Loschmidt echo for the gapped and gapless system under a magnetic field quench, for $A = 100$, $h = 0.05$. The gapped system approaches a constant value $e^{-c_uAh^2}$ at long times, while the gapless system exhibits algebraic decay $\sim t^{-c_lAh^2}$ in a manifestation of the Orthogonality Catastrophe. \label{fig:GMagField}}
    \end{figure}

The exact numerical results (for the second-order cumulant expansion) for the gapped and gapless systems subject to a magnetic field quench are plotted in Fig.\ref{fig:GMagField}. At long times we observe that in contrast to the gapped case, the gapless system has algebraically decaying coherence $|G(t)|\sim t^{-Ah^2c_{h1}}$. We see that for larger $Ah^2$ the decay is faster - that is, the coherence diminishes more quickly for larger systems and stronger fields. The lack of exponential decay is itself an interesting feature with respect to other results in this study - as we will demonstrate, adding noise causes the leading behavior of the coherence decay to be exponential rather than algebraic.

%\mvcomment{Not clear to me why. Do you mean to say because it is the less generic feature when compared to the other results. If so I would just foreshadow a bit with something like a statement "as we will see later". Alternatively if you have literature that points out something exciting about the absence of exponential decay that would be even nicer.}

%\mvcomment{I would discuss the result a bit. I.e. what you can learn from it. Even things that may seem trivial to us like pointing out the are dependence etc. can be useful - especially if you find a way to relate it to what might be interesting to an experimentalist.}

\subsection{Loschmidt echo under the impurity quench}

Next, we study the impurity quench given in Eq.\eqref{eq:imp}. We begin by once again studying the gapped system.

\subsubsection{Gapped system}

The gapped system under impurity quench is analyzed in exactly the same way as the gapped system under a magnetic field quench. In fact, the time-dependent part of Eq.(\ref{fullimpecho}) for the Loschmidt echo has the same stationary phase result as was found for the magnetic field, up to a constant $4A$ as shown below. %This is because $|u(\vec{k}_0)|^2=1$ at each stationary point. 
We find 
\begin{dmath}
    |G_l(t)| \approx e^{-c_l\lambda^2} \exp{\left[ \frac{\lambda^2}{\pi t} \left(-\frac{11^{3/2}}{144}\sin{(2\sqrt{\frac{3}{11}}t)}\right)\right]}\times \exp{\left[\frac{\lambda^2}{\pi t} \left( \sqrt{\frac{11}{3}}\frac{11}{216} \cos{(6\sqrt{\frac{3}{11}}t)}\right) \right]} \times \exp{\left[\frac{\lambda^2}{\pi t} \left( \frac{11\sqrt{55}}{3600}\sin{(10\sqrt{\frac{3}{11}}t)}\right)\right]}.
\end{dmath}
The primary difference between the uniform magnetic field quench and the impurity quench for the gapped phase is in the value $e^{-c_l \lambda^2}$ to which the Loschmidt echo converges in the limit $t\to \infty$:
$$c_l = \int_{-\pi}^{\pi}\frac{d^2k}{(2\pi)^2} \frac{1}{E^2_k}.$$

For parameter choice $\vect J=\sqrt{3/11}(1,1,3)$ we have $c_l = 0.136649$. This means that for the gapped system under an impurity quench, the size of $|G_l(\infty)| = e^{-0.136649\lambda^2}$ is determined fully by the perturbation strength $\lambda$, with finite overlap with the Kitaev ground state as $t \to \infty$. As in the case of the magnetic field, the Orthogonality Catastrophe does not manifest for the impurity quench. Note that $G_l$ is not suppressed by system size in the same way that $G_u$ is. To a first approximation we can also see that the decay should be weaker when the system's gap is larger, because for a larger gap $c_l$ should be smaller. We can see this by imagining that $E_k$ in the integrand of $c_l$ is a constant: when $E_k$ is larger (corresponding to a larger gap) the integral is smaller.

\subsubsection{Gapless system}

For the gapless system with $\vect J = (1,1,1)$, we follow the same procedure as in the gapless system under a magnetic field quench we discussed above. We first analyze the stationary phase result for $\frac{d^2I}{dt^2}$ in $|G_l(t)| \approx \exp{[\lambda^2I(t)]}$, where for the impurity we found in Eq.\eqref{fullimpecho}
$$I(t) = \int_{-\pi}^{\pi}\frac{d^2k}{(2\pi)^2} \frac{\cos{(E_kt)}-1}{E^2_k}.$$

%\mvcomment{I would make clearer that you are actually looking at the Loschmidt echo in this section. Because currently it can be misunderstood as you saying that you look at $\frac{d^2I}{dt^2}$. A simple way to work around this is to just reference the equation with the integral, then mention that it is solved with the techniques of the appendix you reference and/or your mathematical method section. This avoids confusion and should make the writing more clear. You could start something like:" \textit{In this section we contrast the results for the impurity quench for a gapped system with results for a gapless system. The integrals in [reference equation] were evaluated according to the methods outlined in [reference1 and reference 2]. Particularly we find that the Loschmidt echo for a representative parameter choice of $\vect J=\sqrt{3/11}(1,1,3)$ is given as:}"} 

%WR: in the below I restate some of the techniques, not sure if that is helpful or cumbersome.

Once again the reason for beginning with a stationary phase result for the second derivative is that the second derivative has an integrand with no singularities. Upon finding this result we employ the rest of the semi-analytical method detailed in Appendix \ref{app:C}. 
%\mvcomment{Calling the integration constants higher order terms could be misleading because they are the actually dominant terms and usually sub-dominant terms are called higher order. Maybe a good way to put it could be(and a similar statement might be useful to clarify the result from the earlier section): Obviously, integrating $d^2I/dt^2$ twice leads to undetermined integration constants. Through an analysis of singular points of $1/E_k^2$ that we detail in Appendix D we find which of these integration "constants" that arise  dominate the long time behaviour of $I(t)$. }
Following this process gives the long-time result

\begin{equation}
   |G_l(t)| \approx \exp{\left[ \lambda^2(-c_{l0} -c_{l1} \log{t} + \mathcal{O}(1/t))\right]} ,
\end{equation}
where $c_{l0} = 0.219$, and $c_{l1} = 0.091$. We see that the decay much like the gapless case with a constant magnetic field quench is algebraic to leading order. However, in contrast to the magnetic field case, the decay under an impurity quench does not depend on system size, but only on the impurity strength and the value of $c_{l1}$. 

\begin{figure}[htp]
    \centering
    \includegraphics[width=8.5cm]{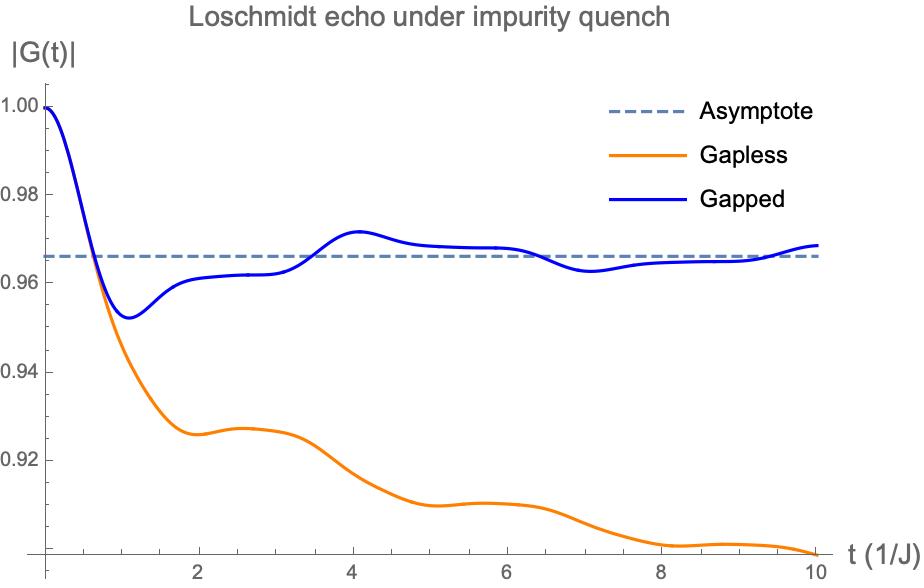}

    \caption{Numerical plots of the Loschmidt echo for gapped ($\vect{J} = \sqrt{3/11}(1,1,3)$) and gapless ($\vect{J} = (1,1,1)$) system under an impurity quench, for $\lambda = 0.5$. The gapped system approaches a constant value $e^{-c_l\lambda^2}$ at long times, while the gapless system exhibits exponential decay in a manifestation of the Orthogonality Catastrophe.  \label{fig:GImp}}
    \end{figure}

To demonstrate the above behavior the Loschmidt echo under an impurity quench is compared for the gapped and gapless systems in Fig.\ref{fig:GImp}. The figure clearly shows that the gapped case reaches a constant asymptotic limit and the gapless decays algebraically in the limit $t\to\infty$.

\begin{figure}[htp]
    \centering
    \includegraphics[width=8.5cm]{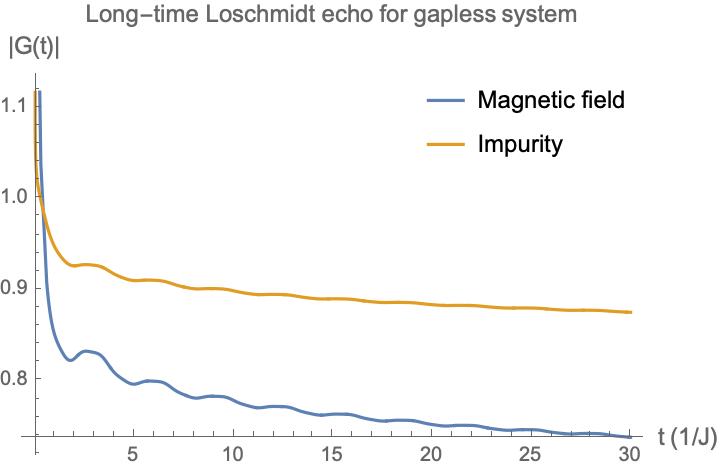}

    \caption{Long-time plots of asymptotic results for Loschmidt echo for the gapless system ($\vect{J} = (1,1,1)$) under magnetic field and impurity quenches, for $\lambda = 0.5$, $h=0.05$, and $A=100$. Even for relatively small field $h$ and strong impurity $\lambda$, the system under a magnetic field quench decays much faster. The $|G(0)|$ behavior is artifical and is due to the inclusion of higher-order stationary phase results $\mathcal{O}(1/t)$, which are not accurate at small times. These higher-order terms also contribute the oscillatory behavior. \label{fig:AsympGaplessPure}}
    \end{figure}

Furthermore, in Fig.\ref{fig:AsympGaplessPure} we directly compare asymptotic results for the gapless system under magnetic field and impurity quenches. Both display algebraic decay, with the magnetic field-quenched system manifesting a faster decay, as expected due to its the dependence on system size.

\subsection{General result for noiseless quenches in the gapped phase} 
So far we have considered a specific gapped system with $\vect J=\sqrt{3/11}(1,1,3)$. However, for the quenches discussed so far (those without noise) we are able to fully analytically obtain general results for gapped systems, which we discuss in this section. 

Because we can apply the stationary phase approximation Eq.\eqref{spformula} directly for the noiseless quenches applied to the gapped phase, we can find a more general expression for the Loschmidt echo for any arbitrary choice of couplings $\vect J$ - of course with the restriction that they give rise to a gap. For the sake of concreteness and to be consistent with the the choice of couplings $\vect J=\sqrt{3/11}(1,1,3)$ we analyze, we consider the region where $J_z > J_x + J_y$. We find

$$ |G_u(t)| \sim \exp{\left[4Ah^2 \left(-c_u + \frac{\gamma(t)}{4\pi \sqrt{J_x J_y J_z}t}\right)\right]}, $$
$$|G_l(t)| \sim \exp{\left[\lambda^2 \left(-c_l + \frac{\gamma(t)}{4\pi \sqrt{J_x J_y J_z}t}\right)\right]},$$
\begin{dmath}
    \gamma(t) = -\frac{\sin{(\Delta E t)}}{\sqrt{2} \Delta E^{3/2}} +\frac{2\cos{(\frac{1}{2} (\Delta E + \omega + \delta)t)}}{(\Delta E +\omega + \delta)^{3/2}}+\frac{2\cos{(\frac{1}{2} (\Delta E + \omega - \delta)t)}}{(\Delta E +\omega - \delta)^{3/2}} +\frac{\sin{(\omega t)}}{\sqrt{2} \omega^{3/2}}.
\end{dmath}

\begin{figure}[htp]
    \centering
    \includegraphics[width=8.5cm]{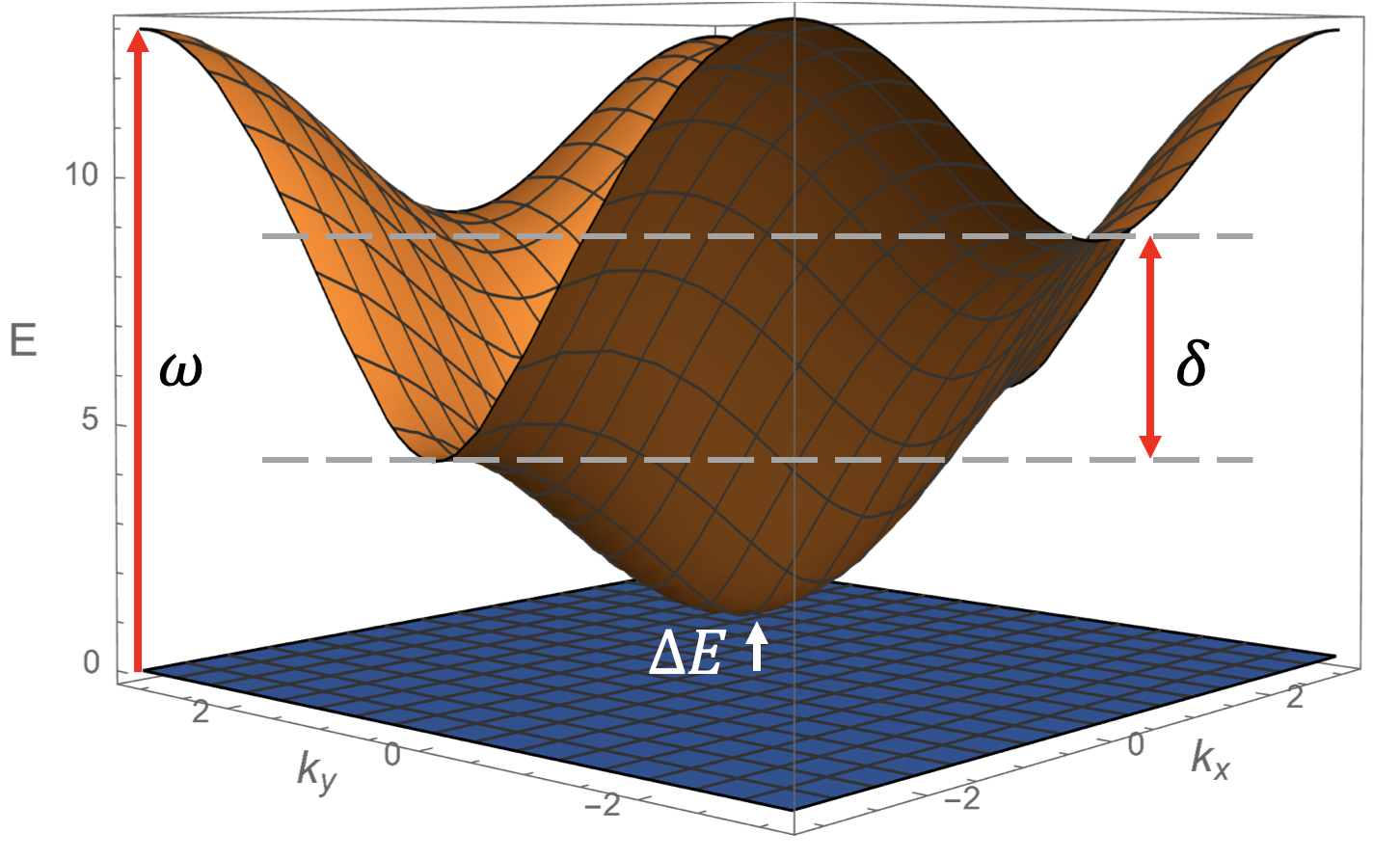}

    \caption{The natural energy scales ($\Delta E$, $\omega$, and $\delta$) of the problem determine the frequencies of oscillation for the Loschmidt echo. $\Delta E$ and $\omega$ are the band minima and maxima respectively, with $\Delta E$ also being half the band gap. $\delta$ is the energy difference between saddle points in the spectrum, and is inherently 2D. \label{fig:2DBandFig}}
    \end{figure}

We find that the result is oscillatory and that there are three natural energy scales of the system that determine the frequencies. $\Delta E = 2(J_z - J_x - J_y)$ is the band minimum (half the band gap), $\omega = 2(J_x + J_y + J_z)$ the band maximum (half the bandwidth), and $\delta = 4(J_x - J_y)$ is the separation between saddle points, as shown in Fig.\ref{fig:2DBandFig}.

   % \mvcomment{I suggest adding a brief discussion of the oscillatory behaviour - that is maybe a brief note that you find partial revivals in particular. If I recall we did not find any very strong revivals so a brief discussion could be better than going into details.}

These energy scales determine the oscillatory behavior of the Loschmidt echo, even at relatively short times. In particular, they determine the frequency at which the Loschmidt echo oscillates around its asymptotic value. We note that oscillations are more visible at earlier times because they are multiplied by a factor $1/t$. 

Interestingly, the oscillatory behavior means that the Loschmidt echo will have periodic revivals. The energy scales given above also pick out the timescale on which any such revivals in the coherence happen. We note that because of the $1/t$ factor revivals are most pronounced early in the evolution. Here, after an initial decay below the asymptotic ($t\to\infty$) value the Loschmidt echo recovers to some local maximum that is above the asymptotic value. This behavior is clearly seen in both Fig.\ref{fig:GMagField} and Fig.\ref{fig:GImp}.

\subsection{Fidelity under environmental coupling}

Next, we study the asymptotic behavior of the fidelity under a noisy magnetic field quench, which is equivalent to a specific form of environmental coupling. Explicit expressions for its behavior are found by approximating the integrals in Eq.\eqref{Eq:alpha} for long times. Due to the divergent denominator $1/(E_k - E_{k'})^2$ that appears in some of the integrals it is not possible to directly apply the stationary phase approximation for all contributions to the integral, regardless of whether the system is gapped or gapless. Rather, we follow the procedure in Appendix \ref{app:noise} to evaluate such integrals in both cases.

Applying this methodology to Eq.\eqref{Eq:alpha}, we arrive at

\begin{dmath}
    F_u(t)\sim \exp{\left[(-\alpha Ah^2\kappa + \beta A^2h^4\kappa^2)t\right]} \times \exp{\left[- h^4 A^2\kappa^2(\gamma \log{t} + \delta)\right]} ,
    \label{eq:fidelitynoisy_quench}
\end{dmath}
for both the gapless and gapped systems. Here the coefficients $\alpha$, $\beta$, $\gamma$, $\delta$ determined by the asymptotic methods differ between the gapped $\vect J=\sqrt{\frac{3}{11}}(1,1,3)$ and the gapless $\vect J=(1,1,1)$ case, both of which are given in Tab.\ref{Tab:FidCoef}.

\begin{table*}[t]
 \centering
\begin{tabular}{|c | c | c | c | c |} 
\hline 
$F_u(t)\sim \exp{\left[(-\alpha Ah^2\kappa + \beta A^2h^4\kappa^2)t - h^4 A^2\kappa^2(\gamma \log{t} + \delta)\right]}$  & $\alpha$ & $\beta$ &  $\gamma$ & $\delta$ \\ [1ex]
 \hline
 Gapped & 3.880 & 13.443 & 2.229 & 7.134  \\ [1ex]
 \hline
 Gapless & 3.050 & 5.859 & 0.709 & 2.719 \\
 [1ex] 
 \hline
\end{tabular}

\begin{tabular}{| c | c | c | c |} 
\hline 
$F_l(t) \sim \exp{\left[(-\kappa + \beta\frac{\kappa^2}
{2})t - \gamma \frac{\kappa^2}{2} \log{t} - \delta \right]} $  & $\beta$ &  $\gamma$ & $\delta$ \\ [1ex]
 \hline
 Gapped & 0.890 & 0.139  & 0.447  \\ [1ex]
 \hline
 Gapless  & 0.645 & 0.076 &  0.275\\
 [1ex] 
 \hline
\end{tabular}

\caption{Coefficients arising in the asymptotic form of a second-order cumulant expansion for fidelity $F(t)$ in gapped $\vect J = \sqrt{3/11}(1,1,3)$ and gapless $\vect J = (1,1,1)$ cases and for uniform field (top) and impurity quench (bottom). The dominant contribution comes from the first cumulant, proportional to $\kappa$, and leads to exponential decay in the fidelity. The $\kappa^2$ terms arise from the second cumulant. Notice that while the dependence on physical parameters is different, with $F_u$ notably depending on system size, the general form $F \sim \exp{[-at -b\log{t} - c]}$ is found for all noisy cases studied. \label{Tab:FidCoef}}
\end{table*}

In both cases there is a dependence on system size $A$, noise strength $\kappa$, and magnetic field strenght $h$. For the cumulant expansion to be reasonable we require small perturbations - that is we can assume $\kappa A h^2  \lesssim 1$ to be a small parameter.  

\begin{figure}[htp]
    \centering
    \includegraphics[width=8cm]{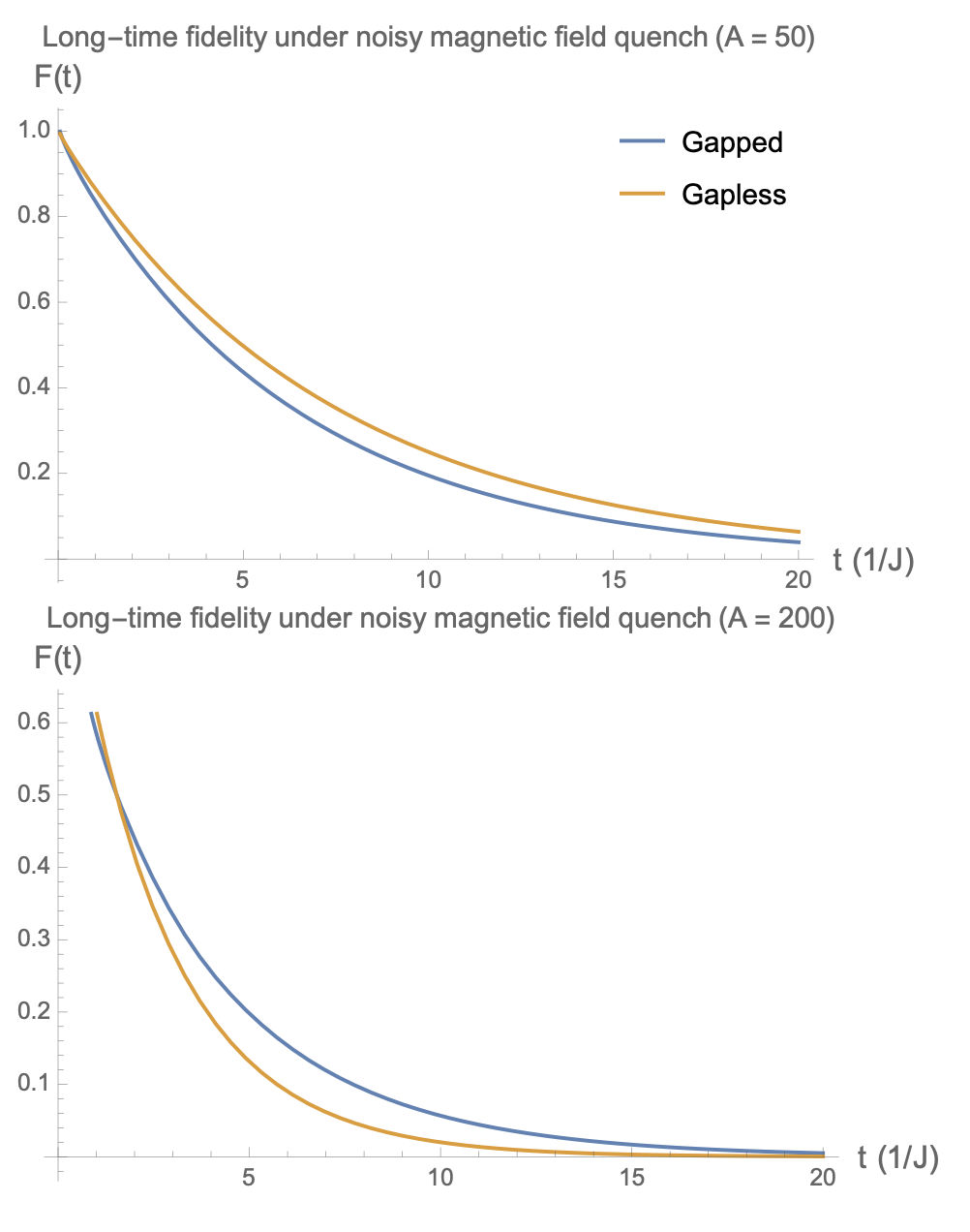}

    \caption{(TOP) Asymptotic behavior of the Uhlmann fidelity for the Kitaev honeycomb ground state with environmental coupling, for $\lambda = 0.1$, $h=0.1$, and $A=50$. The gapped system has $\vect{J} = \sqrt{3/11}(1,1,3)$ while the gapless has $\vect{J} = (1,1,1)$. (BOTTOM) Fidelity for $\lambda = 0.1$, $h=0.1$, and $A=200$. While the fidelity does not meaningfully distinguish between the presence or lack of a gap, it is sensitive to the parameter $\kappa A h^2,$ shown here by varying system size. A larger system (larger $\kappa A h^2$) displays faster decoherence. \label{fig:NoisyFieldFid}}
    \end{figure}

The gapped and gapless decay for any choice of parameters track each other much more closely than in the noiseless cases, where we saw algebraic decay for the gapless system and decay to a non-zero asymptote for the gapped system. We will see that this closeness of the decay between gapped and gapless systems also holds for the noisy impurity, as presented in the next subsection. This suggests that when noisy coupling occurs, the fidelity is not very sensitive to whether the system is gapped or gapless - noteably, the presence of a gap does not result in a meaningful boost to the fidelity. This is due to two structural similarities between the gapped and gapless cases. First, the leading behavior is determined by the first cumulant, which gives exponential decay regardless of whether the system is gapped. The second similarity is more technical and is detailed in Appendix \ref{app:noise}. Here we just provide a brief summary. 

In both the gapped and gapless cases, the leading behavior of the second cumulant at long times is contributed by an integral 
$$\propto \int{\int \frac{d^2k}{(2\pi)^2} \frac{d^2k'}{(2\pi)^2} |u_k|^2|u_{k'}|^2 \frac{1 - \cos{(E_k - E_{k'})t}}{(E_k - E_{k'})^2}}. $$
Here, the denominator $(E_k - E_{k'})^2$ means that there are singularities even for the gapped system, and in the long-time limit it is these singularities that contribute the leading-order behavior. Moreover, contributions from each singular point should have the same functional dependence in time, meaning that we should expect the long-time limit of the integral to have the same functional form regardless of whether the system is gapped or gapless. This behavior is in stark contrast to the noiseless field and impurity case, where we observed pronounced differences between gapped and gapless cases.

Therefore, the functional form of the long-time result is determined by the first cumulant, giving exponential decay, and an important term in the second cumulant. Up to some constants determined by the integration (written as $\alpha$, $\beta$, $\gamma$, $\delta$), the result for the fidelity under a noisy uniform field quench has the same functional dependence on time for both the gapless and gapped systems. While these constants may be somewhat different in each case, the time dependence of the fidelity is to a large extent determined by the parameter $\nu = \kappa A h^2 $, especially because this determines the leading-order exponential decay. Larger values of $\nu$ result in faster decoherence, as seen in Fig.\ref{fig:NoisyFieldFid}. This has implications for the design of quantum devices subject to environmental noise: in the presence of noise, coherence is maintained for longer in a device that is smaller, or more weakly coupled to the environment through better shielding. 

We note also that of all cases studied in this work, the noisy uniform field quench is the only one that leads to any dependence on $A^2$ of the fidelity (to second order in a cumulant expansion). Therefore, large-area quantum devices are most susceptible to this type of outside disturbance.

\subsection{Fidelity under noisy impurity quench}
Lastly, we study the fidelity of the ground state under a local dissipator or a magnetic impurity subject to white Gaussian noise given by Eq.\eqref{eq:noisy_imp_L}. We note that for both the gapped and gapless phases, methods from the preceding section (see Appendix \ref{app:noise} for details) are applicable. These methods can be used to find the long-time behavior of the integral determining the fidelity under a noisy impurity quench, shown in Eq.(\ref{Eq:localfid}). In this case we find that the asymptotic behavior is given by

\begin{equation}
   F_l(t) \sim \exp{\left[(-\kappa + \beta\frac{\kappa^2}
{2})t - \gamma \frac{\kappa^2}{2} \log{t} - \delta \right]}, 
\end{equation}
where the coefficients $\beta$, $\gamma$, $\delta$ differ in the gapped and gapless phase and two specific cases are given in Tab.\ref{Tab:FidCoef}. 

\begin{figure}[htp]
    \centering
    \includegraphics[width=8.5cm]{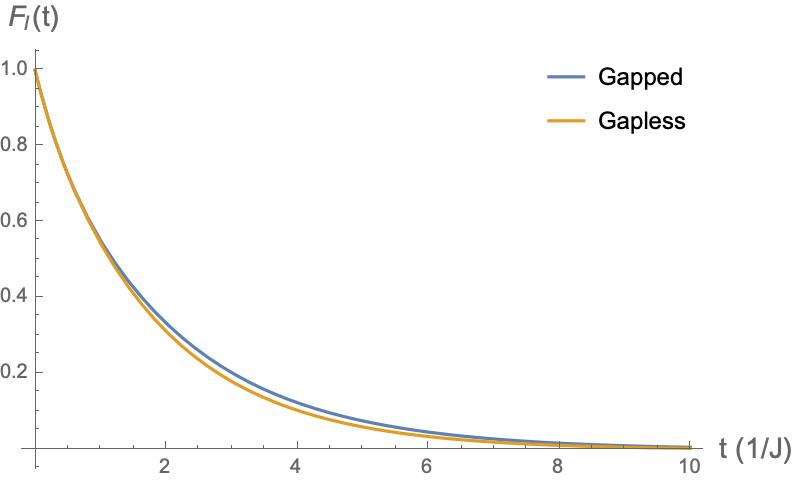}

    \caption{Fidelity for the noisy impurity for the gapped, $\vect{J} = \sqrt{3/11}(1,1,3)$ (blue) and gapless, $\vect{J} = (1,1,1)$ (orange) systems. Even for a relatively large $\kappa$ value, here $\kappa = 0.7$, the distinction between phases is not large. This suggests that in the presence of noise the coherence is not sensitive to the presence of a gap.  \label{fig:LocalFid}}
    \end{figure}

In both cases the linear term leads to exponential decay, which is a hallmark of the Orthogonality Catastrophe, and a logarithmic term gives rise to algebraic decay. While the fidelity for the gapped system decays more slowly than that for the gapless system, this difference is minor, as can be seen in Fig.\ref{fig:LocalFid}. In other words, as was found for environmental coupling, the fidelity is not very sensitive to the presence or lack of a gap, suggesting that this is a general feature of noisy quenches, even when the noise is localized. This  can be explained by the structure of the first cumulant, which is dominant and has the form $\int_0^t ds (\bra{g}L\ket{g}^2 - \bra{g}L^2\ket{g})$ - see Eq.\eqref{eq:firstL}. Because the integrand lacks time dependence, this term will always lead to exponential decay as the leading-order behavior.

We stress the universality of the long-time results for the noisy perturbations, $F(t) \approx Ct^{-b}e^{-a t}$, which applies regardless of the type of perturbation (local or global) or the presence of a gap. This universality arises due to the form of the first cumulant, as well as the nature of the singularities in the integrands for the second cumulant in all noisy cases studied. With denominators $\propto (E_k - E_k')^2$ for both local and global noise, singularities in the integrand contribute to the asymptotic behavior even for a gapped system. While the singular regions cover more of $k$-space for the gapless system, the functional form of the asymptotic contribution is the same for gapped and gapless systems, giving rise to the observed universal behavior. This can be seen by doing a linear expansion in the integrand around a generic singular point, which in general leads to the observed algebraic decay. This general form for the fidelity has also been found for an Ising chain in a noisy magnetic field, highlighting that this form is specific to neither the Kitaev model nor to 2D systems \cite{Marino}.

It is worth noting that the linear expansion of the denominator that produces this form for $F(t)$ is due to the Dirac dispersion around the singular points in the energy band. This is not seen universally for for every gapless choice of parameters $\vect{J}$. Critical points between the gapped and gapless phases instead have quadratic dispersion along one axis near their $E_k = 0$ points. While we do not study these here due to the fine-tuning necessary to find such points physically, this could be an interesting line of future study as it likely would produce distinct asymptotic behavior to the non-critical gapless system.

%\mvcomment{Here we would have the opportunity to point out that there could be specific cases with the denominator not being linear that could lead to other behavior. We could suggest a study of that}

Lastly, we note that just as was the case for the noiseless quenches, here we have found that the noisy impurity quench gives a coherence measure that is independent of system size, while the coherence for a system under noisy field quench does depend on system size - that is, this area-independence occurs regardless of whether the impurity is noisy.

\subsection{Study of excitations}

In the preceding sections we have studied the coherence of the Kitaev honeycomb ground state under various perturbations. However, one might generally wish to know about the coherence of an arbitrary eigenstate of the Kitaev model. The basic excitations of the model are spinons, which we have written as $\gamma_k$, and flux excitations, relating to our $f_r$ fermions. The study of excitations has technological and theoretical relevance; the flux excitations, being related to anyons, are particularly interesting in the context of topological quantum computing. Spinons, on the other hand, have potential relevance to the field of spintronics with coherent spin excitations having the potential to transmit energy through transport effects such as the spin Seebeck effect \cite{Hirobe_2016,PhysRevB.105.115137}. Such spin current responses are more likely to be experimentally observable when the spin excitations themselves are robust in the presence of disturbances, providing a possible experimental probe of quantum spin liquid behavior in real materials. Likewise, spinons with longer coherence times have greater potential in the design of spintronics devices. Because the spinons are readily treated using the techniques already discussed in this paper, we will restrict our attention to their coherence here. 

In particular, we begin with a study of the long-time Loschmidt echo of a state with $m$ discrete spinon excitations $\gamma_k$ (see Eq.\eqref{eq:diagham}) for quenches without noise. States with only spin excitations are particularly simple to analyze within the apparatus we have used in the present work. Their treatment also highlights why considering the ground state coherence is useful even if one is interested in the evolution of excited states under a quench. For instance, it is straightforward to compute to second order in a cumulant expansion the quantity
\begin{equation} \label{eq:excitation}
G^{ex}(t) =  \bra{\{k\}}e^{iH_0t}e^{-i(H_0+V)t}\ket{\{k\}},
\end{equation}
where $\ket{\{k\}}= \prod_{\{k\}}\gamma^{\dagger}_k\ket{g}$ is the excited state with an arbitrary number of modes excited, and $\{k\}$ denotes the set of excited modes. It turns out that for the impurity quench ($V = \lambda \sigma^z_r$), Eq.\eqref{eq:excitation} is identical to the ground state Loschmidt echo, so that $G^{ex}_l(t) = G_l(t)$ within the cumulant expansion. This holds for any number of excitations and for any choice of parameters $\vec{J}$, and tells us that our results for the ground state coherence under an impurity quench already capture the long-time coherence of the spin excitations. 

\begin{figure}[htp]
    \centering
    \includegraphics[width=8.5cm]{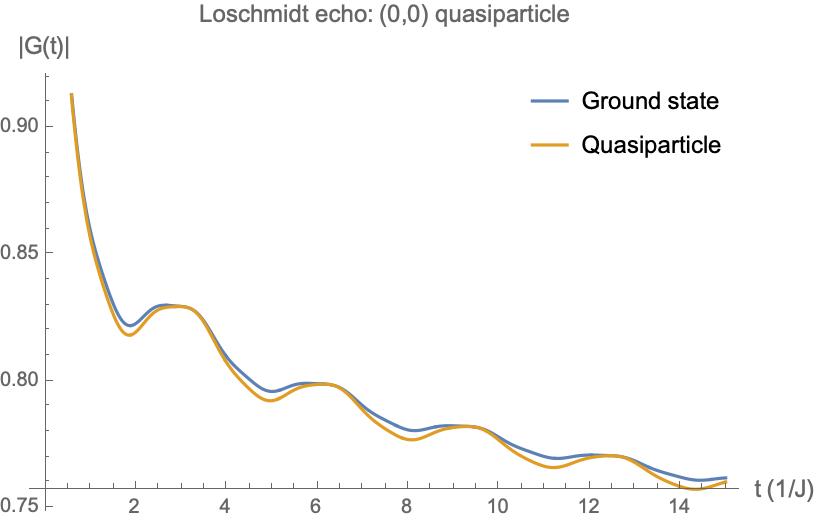}

    \caption{Comparison of the Loschmidt echo of the ground state and an excited state with one spin excitation, for a magnetic field quench applied to the gapless system. As is the case here, most modes do not individually modify the Loschmidt echo significantly compared to the ground state value.  \label{fig:excitation}}
    \end{figure}

We can similarly analyze Eq.\eqref{eq:excitation} for a magnetic field quench with $V = h\sum_r \sigma^z_r$. Contrary to the case of the impurity quench, in this case we find new features arising from the excitations. The spinon coherence is not identical to that of the ground state, but modulates it so that
\begin{dmath}
     |G^{ex}_h(t)| \approx  \exp{\left[4Ah^2\int_{-\pi}^{\pi}\frac{d^2q}{(2\pi)^2} \left[ \cos{(E_qt)} -1\right]\frac{|u_q|^2}{E^2_q}\right]}\times \exp{[4h^2 \sum_{\{k\}}\frac{\epsilon_k}{E_k}(\frac{1-\cos{(E_kt)}}{E^2_k})]}
\end{dmath}
$$=|G_h(t)| \times \exp{[4h^2 \sum_{\{k\}}\frac{\epsilon_k}{E_k}(\frac{1-\cos{(E_kt)}}{E^2_k})]} ,$$
where the sum $\sum_k$ is over the occupied excited modes. 

When we fill regions of the Brillouin zone continuously, the sum becomes an integral over the filled regions, so that 

\begin{equation} \label{eq:exechoh}
     |G^{ex}_h(t)| \approx |G_h(t)| \exp{[4Ah^2 \int_{\{k\}} \frac{d^2k}{(2\pi)^2}\frac{\epsilon_k}{E_k}(\frac{1-\cos{(E_kt)}}{E^2_k})]} .
\end{equation}

Here the first factor is just the ground state result, and now each excitation contributes a factor $\exp{[4h^2 \frac{\epsilon_k}{E_k}(\frac{1-\cos{(E_kt)}}{E^2_k})]}$. While occupied modes in most regions in the Brillouin zone do not modify the ground state Loschmidt echo by much (see Fig. \ref{fig:excitation}), this is not true for regions near Dirac points ($E_k = 0$) in the gapless phase. Occupied modes around the Dirac points make a non-negligible contribution to the Loschmidt echo, strengthening or weakening the coherence considerably.

\begin{figure}[htp]
    \centering
    \includegraphics[width=8.5cm]{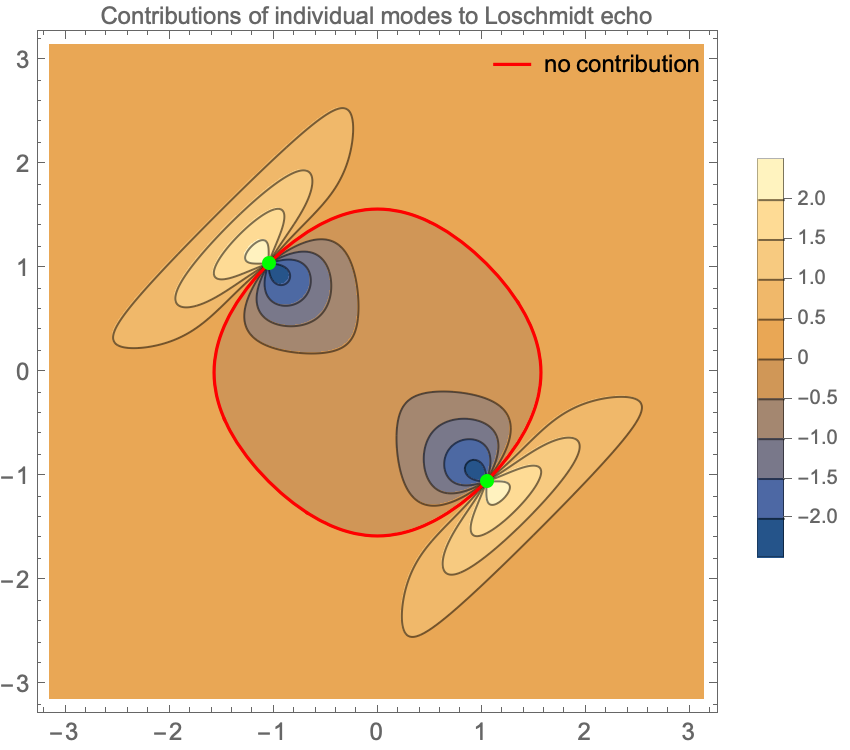}

    \caption{Map of regions giving positive (yellow) and negative (blue) contributions to coherence relative to the ground state coherence for magnetic field quench. This is shown for the specific case of $\vect{J} = (1,1,1)$. Dirac points (green) at $\vect{k} = (\pm \pi/3, \mp \pi/3)$ give singular contributions to the integral in Eq.\eqref{eq:exechoh}, and regions around them contribute most heavily to modifying the Loschmidt echo relative to the ground state result. In particular, the plot is of the integrand in Eq.\eqref{eq:exechoh} at a particular time; the contribution only becomes more heavily localized around the Dirac points at later times. By virtue of being near the Dirac points, the heaviest contributions also occur in the regions of lowest energy. \label{fig:modescont}}
    \end{figure}

By analyzing the sign of the integrand in Eq.\eqref{eq:exechoh}, we can specify the regions of the BZ that contribute positively and negatively to the coherence relative to the ground state. This is shown in Fig. \ref{fig:modescont}, where the interior region (bounded by the red curve in the figure) contains modes that reduce the coherence relative to the ground state calculation, while the exterior region's modes increase it. We see that the greatest contributions lie on either side of the Dirac points at $(\pm \pi/3, \mp \pi/3)$. For example, we can imagine filling some fixed area of the Brillouin zone. Preferentially filling the region outside the red curve and near the Dirac points should give the largest increase to the coherence.

\begin{figure}[htp]
    \centering
    \includegraphics[width=8.5cm]{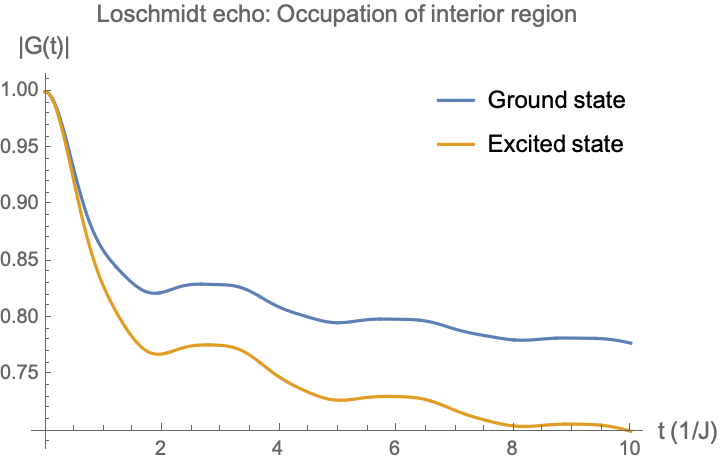}

    \caption{Loschmidt echo for gapless ($\vect{J} = (1,1,1)$) system with interior (negatively contributing) modes fully occupied, under magnetic field quench, for system size $A=100$ and field strength $h = 0.05$. We see that the coherence is diminished relative to the coherence of the ground state under the same quench.  \label{fig:filledint}}
    \end{figure}

\begin{figure}[htp]
    \centering
    \includegraphics[width=8.5cm]{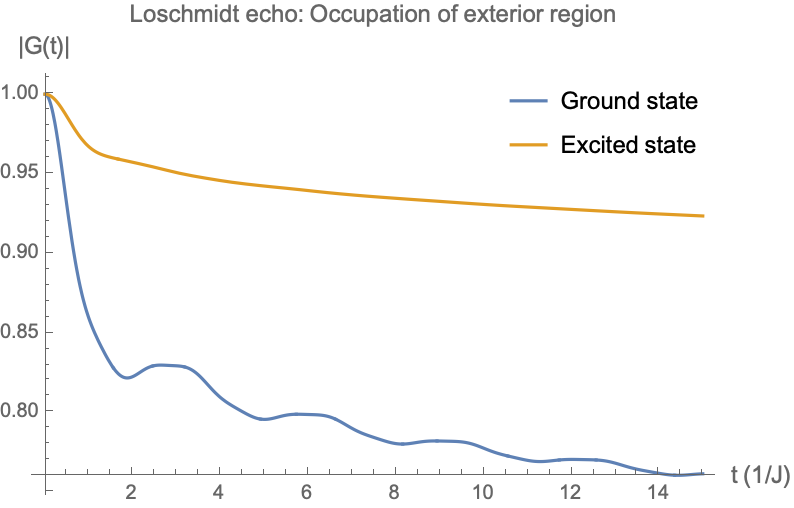}

    \caption{Loschmidt echo for gapless ($\vect{J} = (1,1,1)$) system with exterior (positively contributing) modes fully occupied, under magnetic field quench with $A = 100$ and $h = 0.05$. We see that the coherence is greatly increased relative to the coherence of the ground state under the same quench. \label{fig:filledext}}
    \end{figure}

\begin{figure}[htp]
    \centering
    \includegraphics[width=8.5cm]{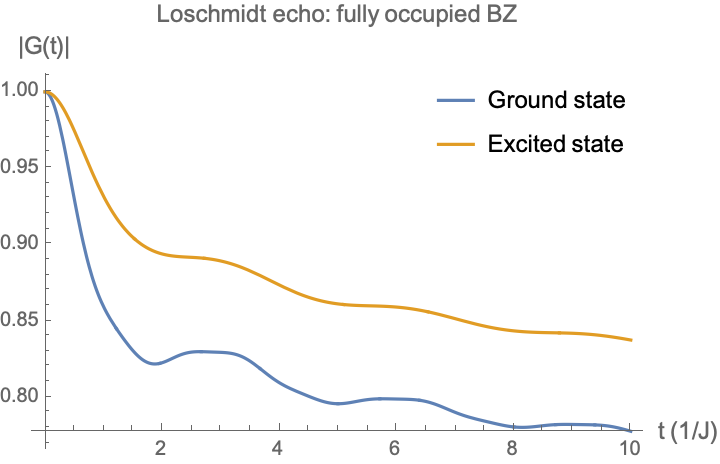}

    \caption{Loschmidt echo for gapless ($\vect{J} = (1,1,1)$) system with Brillouin zone fully occupied, under magnetic field quench, with $A=100$ and $h=0.05$. We see that the coherence is increased relative to the coherence of the ground state under the same quench. \label{fig:filledBZ}}
    \end{figure}

Finally, we explicitly calculate the Loschmidt echo for three filling scenarios: (i) filling only the negatively contributing modes, (ii) only the positively contributing ones, and (iii) the full Brillouin zone. Interestingly, filling the BZ increases the coherence above that of the ground state. These cases are shown in Figs.\ref{fig:filledint}, \ref{fig:filledext}, and \ref{fig:filledBZ} respectively.

We see that completely filling the interior region (case (i), Fig. \ref{fig:filledint}) diminishes the coherence, while filling the exterior region (case (ii), Fig. \ref{fig:filledext}) substantially increases the coherence on this timescale. Interestingly, a fully occupied BZ (case (iii), Fig. \ref{fig:filledBZ}) also increases the coherence above the ground state value. 

Case (ii) is worth lingering on, due to the substantial increase in coherence. While the state with all positively contributing modes filled can decay into one in which interior modes (those in case (i)) are occupied, it is important to note that the greatest negatively contributing modes of the interior region (depicted in Fig. \ref{fig:modescont}) are of similar energy to the greatest positively contributing modes of the exterior region. In fact, the modes in the exterior that contribute most greatly to an increase in the coherence are the ones of lowest-energy, in virtue of their being localized near the Dirac points. They are therefore also the modes that are least likely to decay, for example at finite temperature.

We therefore see that the study of the spin excitations leads to two interesting and distinct conclusions in the cases of the impurity and uniform field quenches. Under an impurity quench, the coherence of a system with any number of occupied spinon modes evolves exactly as that of the ground state. This highlights the generality and importance of the ground state calculation, in that it already captures the relevant physics concerning the Loschmidt echo when spin excitations are of interest. Likewise, in the case of a magnetic field quench the ground state calculation captures the most important behavior when a small number of spinon modes are occupied. However, this is not the case when many states are excited, especially near the Dirac points. In this case the coherence can be strengthened or diminished substantially, suggesting for example that preferential occupation near the Dirac points can significantly increase coherence times.

We close the discussion of spinon excitations by noting that in the gapped phase, the integrand in Eq.\eqref{eq:exechoh} is always positive. Thus, in the gapped phase any excited mode or collection of excited modes grants an increase to the coherence, with the greatest increase coming from regions of lower energy, where the integrand is larger.
    
\section{Outlook} \label{sec:outlook}

We conclude our discussion with an overview of some suggestions for further study. The present work has mostly focused on the long-time coherence of the Kitaev honeycomb ground state, with the previous section making some remarks about the behavior of spin excitations. This leaves two important cases open to study: anyonic excitations, and thermally occupied states. The coherence of anyonic excitations would be particularly interesting to study, given the importance of anyons in topological quantum computing. The formalism presented in this work is apt for such a study: anyons would be built out of flux excitations captured by the $f_r$ particles. Within our formalism, this would likely involve taking $J_z$ to be small, as states with flux excitations do not generally have the property that the second term in Eq.\eqref{eq:vtilde} makes no contribution to the cumulant expansion. However, this would still give one the freedom to sample gapped, gapless, and critical regions in the phase space of the Kitaev model.

Thermally occupied states would require some changes to the present formalism, due essentially to the fact that the cumulant expansion presented here was based on an analysis of a pure quantum state. However, this study would be interesting in capturing the temperature dependence of coherence times, which is clearly relevant to the realization of any technological application.

As mentioned earlier, here we did not study critical points of the model due to the fine-tuning of parameters necessary for the system to exist in such a phase. However, this analysis would likely yield novel results for the long-time behavior of the coherence. This is due to the fact that at these critical points, the spectrum near the zeroes of the energy is no longer Dirac-like, but is instead quadratic along one direction. Because the leading-order asymptotic behavior is decided by the contributions near these energy zeroes, this change in the form of the spectrum is likely to produce a different time dependence from non-critical gapless couplings $\vect{J}$.

\section{Conclusion} \label{sec:conclusion}
   % \mvcomment{\begin{enumerate}
       % \item Summarize what we have done as well as the results
       % \item Importantly speculate a bit how our results can be useful in the future and can help with the development of the currently increasingly robust quantum devices
    %\end{enumerate}}

Here we have studied the long-time coherence of the Kitaev model ground state and spin excitations under various quenches representing possible disturbances to the system. We analyzed four kinds of weak disturbances: an impurity quench representing a local disturbance such as a piece of dust on a device, an applied magnetic field, a noisy impurity that might model a hole in magnetic shielding, and a noisy magnetic field. This final situation of a spatially uniform magnetic field modulated by temporal Gaussian white noise is formally related to the situation in which the Kitaev system is coupled to a magnetic bath, and so those results are applicable to that conceptual picture as well. 

Through a long-time asymptotic analysis of both a gapped and gapless Kitaev system, we found a universal functional form for a coherence measure - the Uhlmann fidelity. In terms of the Uhlmann fidelity, this form is $F(t) \sim Ce^{-\alpha t}t^{-\beta}$ (see Table \ref{Tab:res1}). However, the particular values and parameter-dependence of $C$, $\alpha$, and $\beta$ differ crucially from case to case.
More precisely, we found that the coherence of the gapped system under weak, noiseless disturbances decays asymptotically to a finite value $C$, which is set by the strength of the disturbance $\lambda$ for an impurity $V = \lambda \sigma^z_r$ and the field strength $h$ and system size $A$ for the uniform field quench. The long time asymptotic result of the coherence of the magnetic field-quenched system  behaves like $e^{-Ah^2}$, so that larger systems in stronger fields experience greater decoherence.

The gapless system under noiseless quenches was found to exhibit algebraic but not exponential decay $F(t) \sim Ct^{-\beta}$, i.e. $\alpha = 0$. Again in the case of the magnetic field quench, the coherence decays more quickly for larger systems in stronger fields, where $\beta \sim Ah^2$.

We found that universality in the functional form of the long-time results truly emerges in the case of noisy quenches, where neither $\alpha$ nor $\beta$ is zero, regardless of local or global quench or whether the spectrum has a gap. In fact, we found that the long-time fidelity in these cases is not particularly sensitive to the presence of a gap, and depends much more on the local or global nature of the noisy coupling; again we found that for the case of the noisy field (or bath coupling) a larger system will decohere more rapidly. This recurring dependence on system size for systems coupled to a magnetic field, whether noisy or noiseless, suggests that for device design, system size is important whenever magnetic shielding is not perfect, as smaller systems will feel the effect of the coupling less acutely. 

Finally, we conducted a preliminary study of the long-time coherence of excited states of the Kitaev model. In particular we focused on spin excitations under noiseless quenches. We found in the case of the impurity quench, the ground state result is identical to the coherence of an excited state with any number of spinon modes occupied. This signals that it is the ground state properties that determine the long-time coherence for this case. In the case of the magnetic field quench, the result is slightly more subtle. For small numbers of excited modes, the coherence is not very different from the ground state result. However, by filling larger regions of the Brillouin zone with excitations, the coherence can be drastically increased above the ground state result, suggesting that selective excitation of the system could augment quantum coherence.

In studying the ground-state coherence of the Kitaev honeycomb model, our work gives useful insights into the properties of the ground state that plays host to anyons. Our study has analyzed the robustness of the ground state that acts as a foundation for anyonic excitations, and thus is an important part of the coherence of the anyons in the model. The other relevant part of this story is the coherence of the anyons themselves. Looking forward, the framework used for the present work could be readily applied to a direct study of anyons in the Kitaev model. Such a direct study would be fruitful in the context of topological quantum computing, where anyonic excitations are proposed as potential qubits due to their inherent topological robustness. 

%\mvcomment{Maybe it could be nice to mention how this work also gives useful insights into the properties of the ground state that plays host to anyons. Then tell that it is part of the story but that studying coherence of anyons directly in future work would give the other part of a relevant story.}

\acknowledgments
We gratefully acknowledge funding from NSF grant DMR-2114825. W.R. acknowledges support from the US Department of Defense through the NDSEG Fellowship. M.V. gratefully acknowledges the support provided by the Deanship of Research Oversight and Coordination (DROC) at King Fahd University of Petroleum \& Minerals (KFUPM) for funding this work through start up project No.SR211001.

\bibliographystyle{unsrt}
\bibliography{literature}

%%%%%%%%%%%%%%%%%%%%%%%%%%%%%%%%%%%%%%%%%%%%
%%%%%%%%%%%%%%% APPENDIX %%%%%%%%%%%%%%%%%%%
%%%%%%%%%%%%%%%%%%%%%%%%%%%%%%%%%%%%%%%%%%%%
\onecolumngrid
\appendix

\section{Definition of the $f_r$ fermions}
\label{app:appendixA}

\begin{figure}[htp]
    \centering
    \includegraphics[width=11cm]{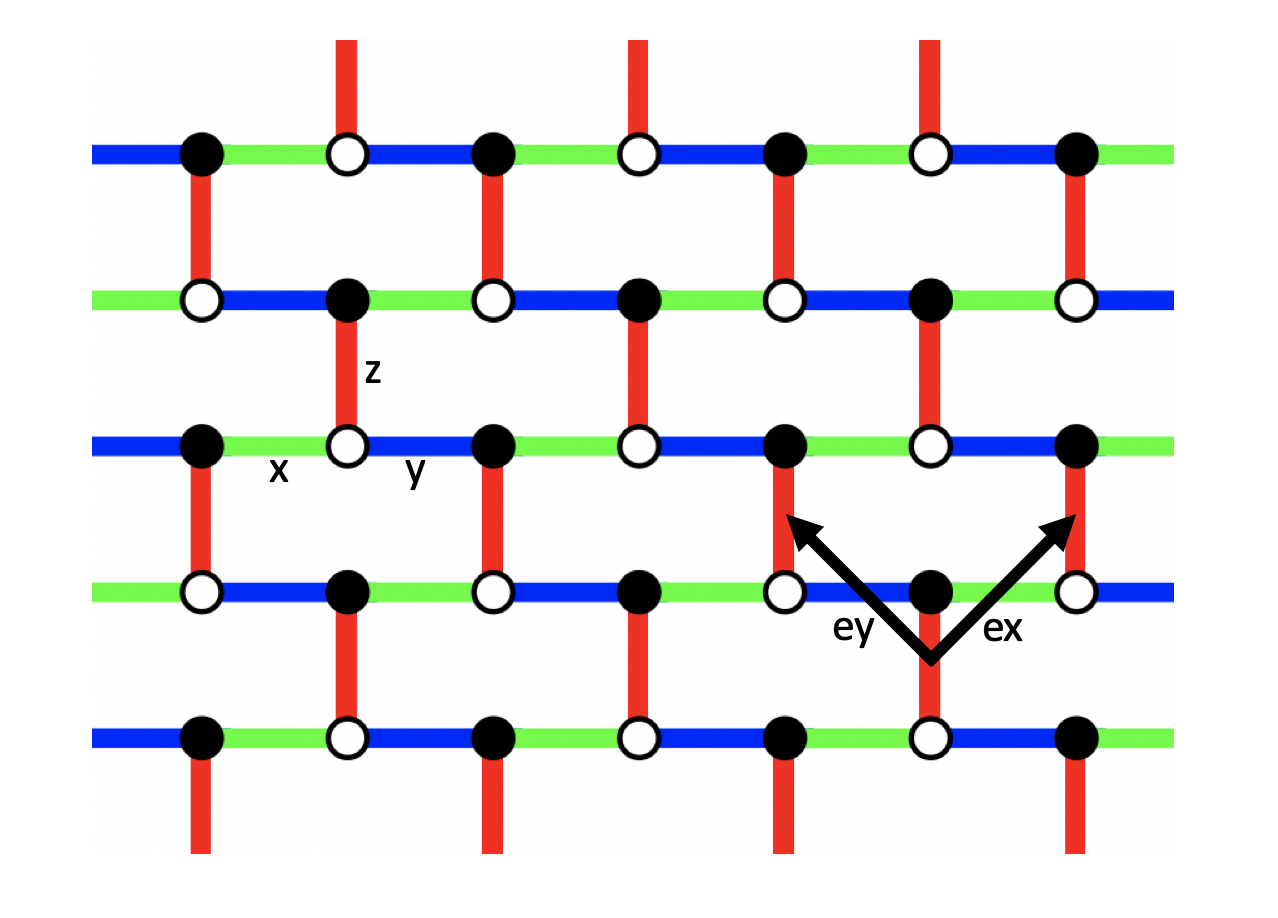}

    \caption{Brick-wall lattice as a deformation of the honeycomb.  \label{fig:bw1}}
    \end{figure}

The transformation between spins and $f_r$ fermions proceeds analogously to that taking spins to $d_r$ in Chen and Nussinov (2008) \cite{Chen_2008}. To clarify the former, the latter is recapitulated in this Appendix. 
The transformation from spins to quasiparticles proceeds in several steps. First, a Jordan-Wigner transformation is performed using the relations 
\begin{equation} \label{eq:A1}
    \sigma^{+}_i = \prod_{j<i}\left( \sigma^z_j\right) c^{\dagger}_i,
\end{equation}
\begin{equation}
    \sigma^{z}_i = 2 c^{\dagger}_i c_i, - 1
\end{equation}
where $\sigma^+_i$ is the spin raising operator at site $i$, and the product of Pauli matrices in Eq.\eqref{eq:A1} is taken along a unique contour in the honeycomb up to and not including site $i$. This contour can be imagined by deforming the honeycomb into a brick-wall lattice, as in Fig.\ref{fig:bw1}. 

Next, the Jordan-Wigner fermions are transformed to a Majorana representation, according to

$$iA_w = c_w - c^{\dagger}_w, \; \; \;\;\; \;  B_w = c_w + c^{\dagger}_w,$$
\begin{equation}
    iB_b = c_b - c^{\dagger}_b, \; \; \;\;\; \; A_b = c_b + c^{\dagger}_b,
\end{equation}
where the operator indices designate whether the operator is for a black or white site. So far the full Hilbert space of the system is accessible by these operators; we have not ignored any degrees of freedom and we have not added any auxiliary non-physical degrees of freedom, either. However, next Chen and Nussinov apply a transformation back to fermions, this time living on the $z$-bonds (vertical rungs of the brick-wall) by mixing the Majoranas of a $z$-bond's black and white sites. This is given by: 
\begin{equation} \label{dferms}
    d_r = \frac{1}{2} \left(A_w + iA_b\right),
\end{equation}
where $r$ labels the $z$-bond, and $b$ and $w$ are the black and white sites connected by that $z$-bond. One can check that Eq.\eqref{dferms} satisfies fermionic anticommutation relations. Notice the lack of analogous fermions defined by mixing of the $B$ Majoranas. Following these transformations, the Kitaev Hamiltonian Eq.\eqref{eq:1} takes the form
\begin{equation}
      H = \sum_r J_x(d^{\dagger}_r + d_r)(d^{\dagger}_{r+ex} - d_{r+ex}) + J_y(d^{\dagger}_r + d_r)(d^{\dagger}_{r+ey} - d_{r+ey}) + J_z\alpha_r(2d^{\dagger}_r d_r - 1),
\end{equation}
where $\alpha_r = iB_b B_w,$ again with $r$ labeling the z-bond connecting $b$ and $w$  \cite{Chen_2008}. A simplification of the model is made by recognizing that $\alpha_r$ is a conserved quantity in the Kitaev model, and will take the eigenvalue 1 in the sector of Hilbert space containing the ground state $\ket{g}$ \cite{Chen_2008}. Thus, the model can be diagonalized within this sector by setting $\alpha_r = 1$ everywhere and ignoring the $B$ Majorana degrees of freedom.  

However, dynamics will in general access the full Hilbert space, and so we need not only to define the ground state with reference to the $d$ fermions containing information about $A$, but also with respect to the $B$ operators. To this end, we can define an analogous transformation that mixes $B$ Majoranas on a $z$-bond and gives us a second spinless fermion flavor living on the $z$-bonds of the lattice:
\begin{equation}
     f_r = \frac{1}{2}\left( B_b - iB_w \right).
\end{equation}

This particular choice leads to a representation in which there are no $f$ fermions in the ground state, because $\alpha_r\ket{g} = (1 - 2f^{\dagger}_r f_r)\ket{g} = \ket{g}.$ Because the perturbations we apply to the ground state will involve the $B$ and therefore $f$ degree of freedom, it is important to know how this degree of freedom acts on the ground state, i.e. \begin{equation}
    f_r\ket{g} = 0.
    \label{eq:zero_gstfr}
\end{equation}

\section{Cumulant expansion in interaction picture} \label{ApExp}

Working in the interaction picture, we have expressions for the Loschmidt echo and Uhlmann fidelity that look like
\begin{equation} \label{Iecho}
    G(t) = \bra{g} Te^{-i\int^t_0 dt' \tilde{V}(t')}\ket{g},
\end{equation}
\begin{equation} \label{Ifid}
    F(t) = (\rho_0| Te^{\int^t_0 ds \tilde{\mathscr{L}_I}(s)}|\rho_0).
\end{equation}

In this Appendix we first review the form of the time evolution operator in the interaction picture, and then discuss the cumulant expansions of equations like Eq.(\ref{Iecho}) and Eq.(\ref{Ifid}). Because these equations are structurally identical, we opt to directly discuss $F(t)$, because the superoperator formalism is less ubiquitous. 

The fidelity is expressed in the interaction picture as
$$F(t) = (\rho_0|\rho_I(t)) \equiv (\rho_0|U_I(t)|\rho_0),$$
where $|\rho_I(0)) = |\rho_0)$. We need to find an expression for $U_I(t)$, the time-evolution operator in the interaction picture. Letting $U(t) = e^{-iH_1t}$, we have $\rho_I(t) = U^{\dagger}(t)\rho(t)U(t)$. Recall the QME:
\begin{dmath} \label{ApQME}
    \frac{d\rho}{dt} =  -i[H_1, \rho(t)] +  \tilde{\mathscr{L}}[\rho(t)].
\end{dmath}
Differentiating $\rho_I$ we find that in the interaction picture Eq.(\ref{ApQME}) takes the form
\begin{dmath} \label{ApQME}
    \frac{d\rho_I}{dt} = \tilde{\mathscr{L}_I}[\rho_I(t)],
\end{dmath}
where $\tilde{\mathscr{L}_I}[\rho_I(t)] = -i\left[V_I(t),\rho_I(t)\right]+\kappa\mathscr{L}_I(t)[\rho_I(t)]$ and $\mathscr{L}_I$ is generated by the Lindbladian operator $L_I = U^{\dagger}LU$.
In the superoperator formalism this translates to 
\begin{equation} \label{IntQME}
    \frac{d}{dt}|\rho_I(t)) = \tilde{\mathscr{L}_I}|\rho_I(t)).
\end{equation}

We now ask about the form of the time-evolution superoperator $U_I$. It must operate on a density operator so that $U_I(t)|\rho_I(0)) = |\rho_I(t)).$ Substituting into Eq.(\ref{IntQME}) we have
$$ \frac{dU_I}{dt}|\rho_I(0)) = \tilde{\mathscr{L}_I}U_I|\rho_I(0))$$
which must hold for arbitrary choice of the initial state $|\rho_I(0))$. We therefore arrive at a differential equation for the time-evolution operator itself:
\begin{equation}
    \frac{dU_I}{dt} = \tilde{\mathscr{L}_I}U_I.
\end{equation}
Integrating from $0$ to $t$ gives
\begin{equation}
    \int^t_0 ds \frac{dU_I}{ds} = U_I(t)-1= \int^t_0 ds\tilde{\mathscr{L}_I}(s)U_I(s),
\end{equation}
where we have used the boundary condition $U_I(0) = 1$, which needs to be fulfilled for any sensible time evolution. This naturally leads to a Dyson equation for $U_I$ when we iterate the expression $U_I(s)=1 + \int^s_0 ds'\tilde{\mathscr{L}_I}(s')U(s')$, which we have just found
$$U_I(t) = 1 + \int^t_0 ds\tilde{\mathscr{L}_I}(s) + \int^t_0 \int^s_0 ds ds'\tilde{\mathscr{L}_I}(s)\tilde{\mathscr{L}_I}(s') + ... = 1 + \int^t_0 dsT[\tilde{\mathscr{L}_I}(s)] + \frac{1}{2!}\int^t_0 \int^t_0 ds ds'T[\tilde{\mathscr{L}_I}(s)\tilde{\mathscr{L}_I}(s')] + ...$$
where $T[]$ is the time-ordering operator. The second equality can be seen from 
\begin{dmath}
    \int^t_0 \int^s_0 ds ds'\tilde{\mathscr{L}_I}(s)\tilde{\mathscr{L}_I}(s') = \frac{1}{2}\int^t_0 \int^s_0 ds ds'\tilde{\mathscr{L}_I}(s)\tilde{\mathscr{L}_I}(s') + \frac{1}{2}\int^t_0 \int^s_0 ds ds'\tilde{\mathscr{L}_I}(s)\tilde{\mathscr{L}_I}(s') = \frac{1}{2}\int^t_0 \int^t_0 ds ds'\Theta(s-s')\tilde{\mathscr{L}_I}(s)\tilde{\mathscr{L}_I}(s') + \frac{1}{2}\int^t_0 \int^t_0 ds ds'\Theta(s'-s)\tilde{\mathscr{L}_I}(s)\tilde{\mathscr{L}_I}(s') \equiv \frac{1}{2}\int^t_0 \int^t_0 ds ds'T[\tilde{\mathscr{L}_I}(s)\tilde{\mathscr{L}_I}(s')].
\end{dmath}

The introduction of the time-evolution operator therefore allows us to rewrite the Dyson series as a time-ordered exponential, so we arrive at
\begin{equation}
    U_I(t) = Te^{\int^t_0 ds\tilde{\mathscr{L}_I}(s)},
\end{equation}
which is how we can write the fidelity in the form of Eq.\eqref{Ifid}. The derivation of the time-evolution operator for kets in the interaction picture, which gives Eq.\eqref{Iecho}, proceeds in a completely analogous way. 
From here we perform a cumulant expansion on the fidelity - that is we employ a partial resummation that assumes exponential behaviour
$$F(t) = (\rho_0|U_I(t)|\rho_0) = e^{\gamma(t)},$$
so that $\log(F(t)) = \gamma(t).$ We then expand $\log(F(t))$ to second order in $\tilde{\mathscr{L}_I}(s)$ using $\log(1+x)\approx x - \frac{x^2}{2}$, which  gives
$$\gamma(t) = \log(1 + \int^t_0 dsT[(\rho_0|\tilde{\mathscr{L}_I}(s)|\rho_0)] + \frac{1}{2}\int^t_0 \int^t_0 ds ds'T[(\rho_0|\tilde{\mathscr{L}_I}(s)\tilde{\mathscr{L}_I}(s')|\rho_0)] + ...)$$
$$\approx \int^t_0 dsT[(\rho_0|\tilde{\mathscr{L}_I}(s)|\rho_0)] + \frac{1}{2}\int^t_0 \int^t_0 ds ds'T[(\rho_0|\tilde{\mathscr{L}_I}(s)\tilde{\mathscr{L}_I}(s')|\rho_0)] - \frac{1}{2}\int^t_0 \int^t_0 ds ds'(\rho_0|\tilde{\mathscr{L}_I}(s)|\rho_0)(\rho_0|\tilde{\mathscr{L}_I}(s')|\rho_0).$$

Using short hand notation $(\rho_0|\tilde{\mathscr{L}_I}(s)|\rho_0) \equiv (\tilde{\mathscr{L}}_I(s))_0$ and $(T[\tilde{\mathscr{L}}_I(s) \tilde{\mathscr{L}}_I(s')])^c_0 = T[(\rho_0|\tilde{\mathscr{L}_I}(s)\tilde{\mathscr{L}_I}(s')|\rho_0)] - (\rho_0|\tilde{\mathscr{L}_I}(s)|\rho_0)(\rho_0|\tilde{\mathscr{L}_I}(s')|\rho_0)$ for so-called cumulants we can write 
\begin{equation}
    F(t)\approx e^{\int^t_0 ds(\tilde{\mathscr{L}_I}(s))_0 + \frac{1}{2}\int^t_0 \int^t_0 ds ds'(T[\tilde{\mathscr{L}_I}(s)\tilde{\mathscr{L}_I}(s')])^c_0},
\end{equation}
where the redundant time-ordering has been removed from the first cumulant. Once again, the procedure for the cumulant expansion of $G(t)$ is exactly analogous to the expansion for $F(t)$; one only needs to replace the Lindbladian superoperators with the Hamiltonian perturbation $\tilde{V}_I$, taking care that an extra $-i$ appears in the Dyson equation in this case.

\section{Calculation of cumulants} \label{B}
In this appendix we will discuss some of the technical details that would obfuscate the main text but are neverthless useful for computing cumulants in the cases of the different quenches we consider. 

\subsection{Magnetic field}

We first start with the magnetic field quench, where we wish to compute \label{c1}
\begin{equation} \label{echoapp}
    G(t) \approx e^{-i\int^t_0 ds \left<\tilde{V}(s)\right>^c_0}e^{-\frac{1}{2}\int^t_0 \int^t_0 ds ds' \left<T\tilde{V}(s)\tilde{V}(s')\right>^c_0}.
\end{equation}

We begin with the first cumulant. In the fermionic representation we have $ V = 2h \sum_r \left(d_rf_r + f^{\dagger}_rd^{\dagger}_r \right),$ and $\tilde{V} =V -2J_z \sum_r f^{\dagger}_rf_r(2d^{\dagger}_rd_r - 1).$ Because the ground state $\ket{g}$ is the vacuum of the $f$ fermions, this is easily computed:
$$\left<\tilde{V}(s)\right>^c_0 = \bra{g}e^{iH_1s}\tilde{V}e^{-iH_1s}\ket{g} = \bra{g}\tilde{V}\ket{g} = \bra{g}V\ket{g}.$$
The final equality follows from Eq.\eqref{eq:zero_gstfr}, which is $f_r\ket{g} = 0.$ From this same relation we note that $\bra{g}d_rf_r\ket{g} = \bra{g}f^{\dagger}_rd^{\dagger}_r\ket{g} = 0.$ Therefore, we find that 
\begin{equation}
    \left<\tilde{V}(s)\right>^c_0 = 0.
\end{equation}

There is therefore no disconnected piece in the second cumulant, so that
$$\left<T\tilde{V}(s)\tilde{V}(s')\right>^c_0 = \left<T\tilde{V}(s)\tilde{V}(s')\right>_0 = T\bra{g}\tilde{V}(s)\tilde{V}(s')\ket{g}.$$
The goal is then to evaluate $T\bra{g}\tilde{V}(s)\tilde{V}(s')\ket{g}$. Leaving the time-ordering alone for now, we have
\begin{equation}
    \bra{g}\tilde{V}(s)\tilde{V}(s')\ket{g} = \bra{g}e^{iH_1s}\tilde{V}e^{-iH_1s}e^{iH_1s'}\tilde{V}e^{-iH_1s'}\ket{g} = \bra{g}Ve^{-iH_1\Delta s}V\ket{g},
\end{equation}
where $\Delta s = s - s'$. The final equality follows from the fact that $f_r\ket{g} = \bra{g}f^{\dagger}_r = 0$ and $\left[f^{\dagger}_r f_r, d^{\dagger}_r d_r \right] = 0$. This means that the $f$-dependent term from $H_0$ contributes to neither the first nor second cumulant. Continuing to simplify we have:
\begin{dmath}
    {\bra{g}Ve^{-iH_1\Delta s}V\ket{g}} = 2h\sum_r{\bra{g}Ve^{-iH_1\Delta s}f^{\dagger}_rd^{\dagger}_r\ket{g}} = {\frac{2h}{\sqrt{N}}\sum_r\sum_k e^{-i\vec{k}\cdot \vec{r}}e^{-iE_k\Delta s}u_k\bra{g}Vf^{\dagger}_r\gamma^{\dagger}_k\ket{g}},
\end{dmath}
where we have used a Fourier transform $d^{\dagger}_r = \frac{1}{\sqrt{N}}\sum_k e^{-i\vec{k} \cdot \vec{r}}d^{\dagger}_k$, and the transformation to quasiparticles of $H_1$ is given by $ d_k = u^*_k\gamma_k + v_k \gamma^{\dagger}_{-k}.$ Applying the time evolution operator then gives $e^{-iH_1\Delta s}\gamma^{\dagger}_k\ket{g} = e^{-iE_k\Delta s}\gamma^{\dagger}_k\ket{g}$. Expanding the remaining $V$ in the fermion representation and enacting another Fourier transform and quasiparticle transformation gives
\begin{dmath}
    {\bra{g}Ve^{-iH_1\Delta s}V\ket{g}} =  4h^2 \sum_k e^{-iE_k\Delta s} |u_k|^2.
\end{dmath}

Finally, reintroducing the factor $-\frac{1}{2}$ from the cumulant and computing the time-ordered integral in Eq.(\ref{c1}) gives
\begin{equation}
\gamma(t)=-\frac{1}{2}\int^t_0 \int^t_0 ds ds' \left< T\tilde{V}(s)\tilde{V}(s')\right> ^c_0 = 4ih^2\sum_k \frac{1}{E_k} \left(t - \frac{\sin{(E_kt)}}{E_k} \right)|u_k|^2  +4h^2 \sum_k \frac{\cos{(E_k t)}-1}{E^2_k} |u_k|^2 
\end{equation}
with $G(t)\approx e^{\gamma(t)}$. Taking the thermodynamic limit $N\to \infty$ takes the $k$-space sums to integrals over the Brillouin zone of the square lattice, which is the result given in Eq.\eqref{fullmagecho} in the main text.

\subsection{Impurity}
The cumulant calculation for the magnetic impurity proceeds very similarly, except that now $ V = \lambda \left(d_rf_r \pm d_rf^{\dagger}_r \mp d^{\dagger}_rf_r - d^{\dagger}_r f^{\dagger}_r\right)$, where the different signs correspond to placing the impurity at a black or white site. Again following from the fact that $f_r \ket{g} = 0$, the first cumulant is zero: 
$$\left<\tilde{V}(s)\right>^c_0 = \bra{g}V\ket{g} = 0.$$
Therefore, we have once more that 
$$\left<T\tilde{V}(s)\tilde{V}(s')\right>^c_0 = \left<T\tilde{V}(s)\tilde{V}(s')\right>_0 = T\bra{g}\tilde{V}(s)\tilde{V}(s')\ket{g},$$
and $ \bra{g}\tilde{V}(s)\tilde{V}(s')\ket{g} = \bra{g}Ve^{-iH_1\Delta s}V\ket{g}.$ Now we proceed with the calculation for the form of $V$ corresponding to the impurity. Following the same steps as above, we arrive at 
\begin{dmath}
    {\bra{g}Ve^{-iH_1\Delta s}V\ket{g}} = \frac{\lambda^2}{N}\sum_k e^{-iE_k \Delta s} \left( \mp v_{-k}u^*_k \mp u_k v^*_{-k} + |v_{k}|^2 +|u_{k}|^2 \right) = \frac{\lambda^2}{N}\sum_k e^{-iE_k \Delta s},
\end{dmath}
where we use the properties of the Bogoliubov transformation from $d_k$ to $\gamma_k$ which tell us that $|v_{k}|^2 +|u_{k}|^2=1$ and $v_{-k}u^*_k$, $u_k v^*_{-k}$ are odd functions of $k$, so that summing them over an even $k$ interval gives zero. The latter means that the Loschmidt echo will be the same regardless of which sublattice the impurity is placed on. Each of these properties can be seen from the requirement that the fermionic Bogoliubov transformation be unitary, and do not require any particular parameterization of $u_k$ and $v_k$. 

All that remains is to take the time ordered integral to obtain the second cumulant: 
\begin{equation}
\gamma(t)=-\frac{1}{2}\int^t_0 \int^t_0 ds ds' \left< T\tilde{V}(s)\tilde{V}(s')\right> ^c_0 = i\frac{\lambda^2}{N}\sum_k \frac{1}{E_k} \left(t - \frac{\sin{(E_kt)}}{E_k} \right)  +\frac{\lambda^2}{N} \sum_k \frac{\cos{(E_k t)}-1}{E^2_k},  
\end{equation}
where we find $G(t)\approx e^{\gamma(t)}$
Once again, taking the thermodynamic limit $N\to \infty$ reproduces Eq.(\ref{fullimpecho}).

\subsection{Noisy cases}
Though the algebraic details for the cases amount to exercises in Wick's Theorem similarly to the above noiseless cases, some general remarks about the first and second cumulant for $F(t)$ are useful. To second order in the cumulant expansion, we have (in terms of superoperator inner products):
\begin{equation}
    F(t) = \exp{\left[\int^t_0 ds (\tilde{\mathscr{L}}_{I}(s))_0 + \frac{1}{2} \int^t_0 \int^t_0 ds ds' (T \tilde{\mathscr{L}}_{I}(s) \tilde{\mathscr{L}}_{I}(s'))^c_0\right]}.
\end{equation}

From the general definition of $\tilde{\mathscr{L}}_{I}(s)$,  we can represent the superoperator innerproducts in terms of the Lindbladian operator $L$ using conventional bra-ket notation. This simply involves converting the inner products to traces via the rule $(A|B) = \Tr{[A^{\dagger}B]}.$ Applying this definition yields 

\begin{equation} \label{eq:firstL}
    (\tilde{\mathscr{L}}_{I}(s))_0 = \bra{g}L\ket{g}^2 - \bra{g}L^2\ket{g},
\end{equation}
for the first cumulant, given that $L$ is a Hermitian operator. This is not time dependent - therefore for any Hermitian operator $L$ the first cumulant will lead to exponential decay as long as $\bra{g}L\ket{g}^2 < \bra{g}L^2\ket{g}$. That is we will find exponential decay term anywhere, where a cumulant expansion is valid. In general linear terms can also arise in the second cumulant, as we found in the present work. For the second cumulant we find (using the simplified notation for the inner product for now)
$$(\tilde{\mathscr{L}}_{I}(s)\tilde{\mathscr{L}}_{I}(s'))_0 - (\tilde{\mathscr{L}}_{I}(s))_0(\tilde{\mathscr{L}}_{I}(s'))_0$$
\begin{equation}
    = |\bra{g}L_I(s)L_I(s')\ket{g}|^2 + \frac{1}{4}\left[\bra{g}L^2_I(s)L^2_I(s')\ket{g} + \bra{g}L^2_I(s')L^2_I(s)\ket{g}\right] - \frac{1}{2}\bra{g}L^2\ket{g}^2.
\end{equation}

Here we have used the property that for the operators we considered, $\bra{g}L\ket{g} = 0.$ The second cumulant will involve more terms if this is not the case. The rest of the calculation amounts to computing expectation values of operators in the interaction picture. Thus, for a many-body problem the calculation becomes an exercise in Wick's Theorem. 

\section{Integral approximation methods} \label{app:C}
All of the integrals that appear in this work can be computed on long timescales by leveraging a stationary phase approximation. Integrals with singularities require more care, but a long-time fit can still be achieved using this paradigm.

\subsection{Direct stationary phase approximation}
Integrals of the form 
$$I(t) = \int^{\pi}_{-\pi} \frac{d^2k}{(2\pi)^2} f(\vec{k})\cos{(\phi(\vec{k}) t)},$$ where $f(\vec{k})$ has no singularities in the Brillouin zone, can be approximated reliably using a stationary phase analysis \cite{Bender}. In the gapless phase, where $E_k \neq 0$, the time-dependent integrals appearing in Eq.\eqref{fullmagecho} and Eq.\eqref{fullimpecho} have this form, so we can apply this procedure directly to those cases when the system is gapped. It is convenient to work with a complex exponential, so we write
\begin{equation}
    I(t) = \operatorname{Re} \int^{\pi}_{-\pi} \frac{d^2k}{(2\pi)^2} f(\vec{k})e^{i\phi(\vec{k}) t}.
\end{equation}
In the limit of large $t$, the exponential oscillates rapidly, which according to the Riemann Lebesgue lemma leads to cancellations between nearby points in $k$-space. Thus, in this limit the dominant contributions to the integral come from small regions around the stationary points. Denoting these points as $\vec{k}_0$ and assuming there is only one stationary point inside the BZ, the stationary phase approximation amounts to the relation 
\begin{equation}
    \int^{\pi}_{-\pi} \frac{d^2k}{(2\pi)^2} f(\vec{k})e^{i\phi(\vec{k}) )} \approx \int^{\vec{k}_0+\vec{\epsilon}}_{\vec{k}_0-\vec{\epsilon}} \frac{d^2k}{(2\pi)^2} f(\vec{k})e^{i\phi(\vec{k}) t},
\end{equation}
with some small separation $\vec{\epsilon}$ from the stationary point. In the end we will need to sum over all stationary points, treating those on the edge of the BZ carefully. For now, we proceed for one stationary point $\vec{k}_0$ in the bulk of the BZ. Assuming $f(\vec{k}_0) \neq 0$, we take $f(\vec{k}) \approx f(\vec{k}_0)$ and expand $\phi(\vec{k})$ to second order around the stationary point, where $\frac{\partial \phi}{\partial k_i} = 0$ at the stationary point. This reduces the problem to a 2D Gaussian integral:
\begin{equation}
     \int^{\vec{k}_0+\vec{\epsilon}}_{\vec{k}_0-\vec{\epsilon}} \frac{d^2k}{(2\pi)^2} f(\vec{k})e^{i\phi(\vec{k}) t} \approx f(\vec{k}_0) e^{i\phi(\vec{k}_0)t}\int^{\vec{\epsilon}}_{-\vec{\epsilon}} \frac{d^2q}{(2\pi)^2}\exp{(\frac{it}{2}(\phi_{xx}(\vec{k}_0)q^2_x + \phi_{xy}(\vec{k}_0)q_x q_y +\phi_{yx}(\vec{k}_0)q_y q_x +\phi_{yy}(\vec{k}_0)q^2_y))},
\end{equation}
where $\vec{q} = \vec{k} - \vec{k}_0$ and $\phi_{ab} = \frac{\partial^2 \phi}{\partial k_a \partial k_b}$.  This can be written more compactly by introducing a Hessian matrix 
$$\phi''\equiv \begin{pmatrix}
\phi_{xx}(\vec{k}_0) & \phi_{xy}(\vec{k}_0) \\
\phi_{yx}(\vec{k}_0) & \phi_{yy}(\vec{k}_0)
\end{pmatrix},$$
so that the integral takes the standard form of a multivariate Gaussian integral
\begin{equation}
     \int^{\vec{k}_0+\vec{\epsilon}}_{\vec{k}_0-\vec{\epsilon}} \frac{d^2k}{(2\pi)^2} f(\vec{k})e^{i\phi(\vec{k}) t} \approx f(\vec{k}_0) e^{i\phi(\vec{k}_0)t}\int^{\vec{\epsilon}}_{-\vec{\epsilon}} \frac{d^2q}{(2\pi)^2}\exp{(\frac{it}{2}(\vec{q}^{\top} \phi'' \vec{q}))}.
\end{equation}
The final approximation is to take the limits of integration to infinity. This is reasonable precisely in situations where we expect the contribution to the integral to fall off rapidly as we move away from the stationary point. This will introduce an error to the approximation that is much smaller than leading order. 
\begin{equation}
      f(\vec{k}_0) e^{i\phi(\vec{k}_0)t}\int^{\vec{\epsilon}}_{-\vec{\epsilon}} \frac{d^2q}{(2\pi)^2}\exp{(\frac{it}{2}(\vec{q}^{\top} \phi'' \vec{q}))} \approx f(\vec{k}_0) e^{i\phi(\vec{k}_0)t}\int^{\infty}_{-\infty} \frac{d^2q}{(2\pi)^2}\exp{(\frac{it}{2}(\vec{q}^{\top} \phi'' \vec{q}))}.
\end{equation}
What remains is just an exercise in multivariate Gaussian integration. The key difference for stationary points on the edge or corner of the BZ is that we will end up with only a portion of the interval in each dimension. For example, a point on the edge might be $\vec{k}_0 = (0,\pi)$. For such a point we do not initially consider a small interval that extends past $\pi$ in the $k_y$ direction. Instead we take
\begin{equation}
    \int^{\pi}_{-\pi} \frac{d^2k}{(2\pi)^2} f(\vec{k})e^{i\phi(\vec{k}) )} \approx \int^{\pi+\epsilon_x}_{\pi-\epsilon_x} \int^{\pi}_{\pi-\epsilon_y} \frac{dk_x dk_y}{(2\pi)^2} f(\vec{k})e^{i\phi(\vec{k}) t} \approx f(\vec{k}_0) e^{i\phi(\vec{k}_0)t}\int^{\infty}_{-\infty}\int^{0}_{-\infty} \frac{d^2q}{(2\pi)^2}\exp{(\frac{it}{2}(\vec{q}^{\top} \phi'' \vec{q}))},
\end{equation}
while for a stationary point at a corner such as $(\pi, \pi)$ we will have 
\begin{equation}
    \int^{\pi}_{-\pi} \frac{d^2k}{(2\pi)^2} f(\vec{k})e^{i\phi(\vec{k}) )} \approx \int^{\pi}_{\pi-\epsilon_x} \int^{\pi}_{\pi-\epsilon_y} \frac{dk_x dk_y}{(2\pi)^2} f(\vec{k})e^{i\phi(\vec{k}) t} \approx f(\vec{k}_0) e^{i\phi(\vec{k}_0)t}\int^{0}_{-\infty}\int^{0}_{-\infty} \frac{d^2q}{(2\pi)^2}\exp{(\frac{it}{2}(\vec{q}^{\top} \phi'' \vec{q}))}.
\end{equation}
Thus we can extend the integrals to $\infty$ in every case, picking up a factor of $1/2$ for stationary points on edges and $1/4$ those at corners, being careful to note that the corners also pick up an overall minus sign.

Computing the multivariate Gaussian integrals leads to the general formula
\begin{equation} \label{spapp}
    \int^{\pi}_{-\pi} \frac{d^2k}{(2\pi)^2} f(\vec{k})\cos{(\phi(\vec{k}) t)} \approx \operatorname{Re} \sum_{k_0} \left(\frac{1}{2}\right)^p (-1)^{\delta_{p-2}}e^{i\phi(\vec{k}_0)t}f(\vec{k}_0)\frac{i}{2\pi t}(\operatorname{det}\phi'')^{-1/2},
\end{equation}
where $p$ is an integer that for stationary phase points at the corners of the BZ is $p = 2$, for stationary phase points at the edges $p=1$, and for points inside the BZ is $p=0$.  Eq.\eqref{spapp} is the starting point for approximating all of the integrals relevant for the current work, regardless of whether the integrand exhibits singularities. That said, the situation is much more subtle when an integral involves singularities, as we must find a way to capture those contributions to the asymptotic behavior. This approach is discussed in the following section.

\subsection{Noiseless integrals in the gapless phase: stationary phase}

In the gapless phase, the $k$-space integrals that enter into 
Eq.(\ref{fullmagecho}) and Eq.(\ref{fullimpecho}) feature singularities where $E_k = 0$. We call the integrals for the noiseless perturbations $I_u$ and $I_l$ where 
\begin{equation}
    I_u(t) = \int \frac{d^2k}{(2\pi)^2} |u_k|^2 \frac{\cos{(E_kt)} - 1}{E^2_k},
\end{equation}
and
\begin{equation}
    I_l(t) = \int \frac{d^2k}{(2\pi)^2}  \frac{\cos{(E_kt)} - 1}{E^2_k}.
\end{equation}
Given the singularities, it should be expected that important long-time contributions to these integrals come not only from regions of stationary phase, but also from regions near these singular points of $1/E_k^2$. Therefore, a stationary phase analysis fails to capture all of the dominant behavior. Indeed, the leading behavior turns out to be logarithmic, while a stationary phase approximation in 2D can only give a sinusoidally modulated $\frac{1}{t}$ dependence. Thus, the leading behavior must not be from regions of stationary phase, but from regions close to singularities. 

To approximate these integrals at long times, we note that the integrands of  $d^2I_u/dt^2$ and $d^2I_l/dt^2$ have no $1/E^2_k$ term and therefore can be captured well by a direct stationary phase approximation. One strategy of finding approximate long-time asymptotic results is therefore to start with an asymptotic expansion of the second time derivatives and successively reproduce $I_u$ and $I_l$ by integration (of course one has to check that the stationary phase approximation remains valid). Therefore, applying Eq.(\ref{spapp}) to the second time derivatives
$$I_u''(t) = -\int\frac{d^2k}{(2\pi)^2} |u_k|^2 \cos{(E_kt)},$$ and
$$I_l''(t) = -\int\frac{d^2k}{(2\pi)^2}  \cos{(E_kt)},$$
we find the long-time behavior to be
$$I_u''(t) \approx -\frac{1}{2\pi t} \left[\cos{(2t)} + \frac{\sqrt{3}}{2} \sin{(6t)} \right],$$ and
$$I_l''(t) \approx -\frac{1}{4\pi t} \left[3\cos{(t)} + \sqrt{3} \sin{(6t)} \right].$$

\begin{figure}[htp]
    \centering
    \includegraphics[width=18cm]{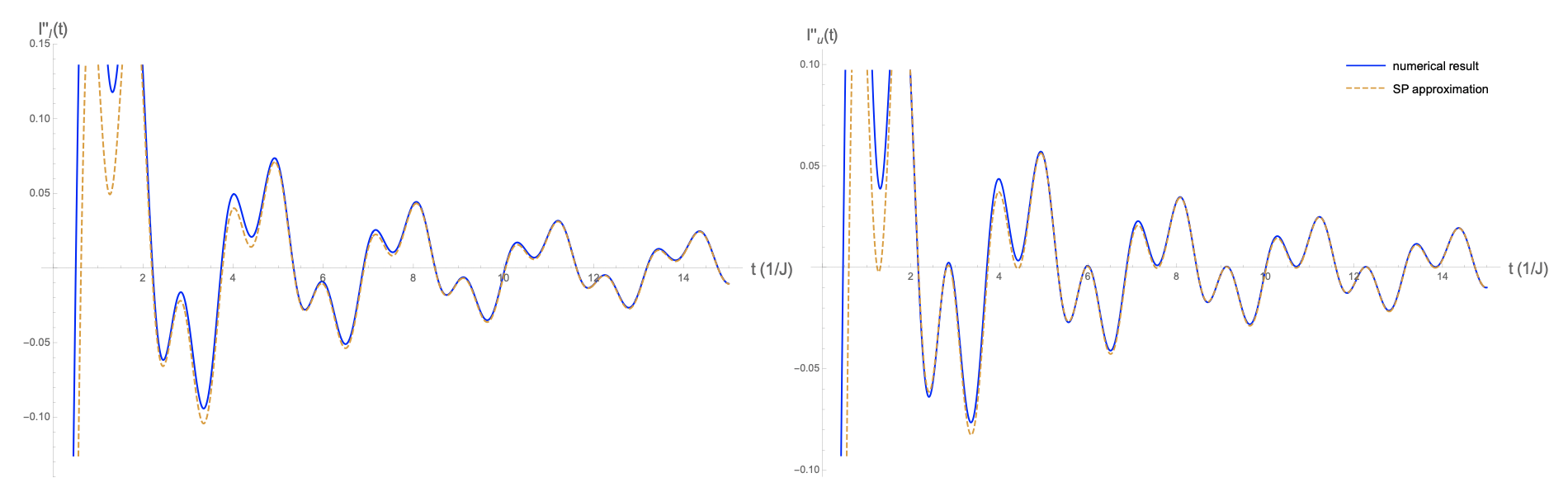}

    \caption{Second time derivative and stationary phase results for $I_l''$ (left) and $I_u''$ (right). Note that while a stationary phase approximation formally requires considering a $t \to \infty$ limit, the approximation captures the behavior well even at intermediate timescales such as those shown here. \label{fig:BothIpp}}
\end{figure}
As seen in Fig.\ref{fig:BothIpp}, in both cases the stationary phase approximation captures the long-time behavior well. Having a good result for the second derivative, we can in principle leverage the fundamental theorem of calculus, integrating twice to obtain information about the first derivative. However, note that the stationary phase result for $I''$ is only valid at long times: therefore when integrating we should choose a large lower bound. That is, 
$$I'(t) \approx I'(t_0) + \int^t_{t_0} I_{approx}''(s) ds \equiv I'_{approx}(t,)$$
$$t > t_0 >> 1,$$
where the inequality ensures that the stationary phase approximation is valid for the entire integration range. Therefore, the inequality ensures that the approximation to $I'(t)$ is valid as an asymptotic limit. Applying the same logic again to the next integral gives
$$I(t) \approx I(t_1) + \int^t_{t_1} I_{approx}'(s) ds, $$
$$t> t_1 > t_0 >> 1.$$

This time the inequality tells us that each subsequent integration pushes the timescale ahead, with the $t_1 > t_0$ being inherited from the first integration. Therefore, we have no guarantee that $I_{approx}''$ and $I_{approx}$ will be valid approximations for the same timescales.
Nevertheless, as outlined above we can approximate $I$ at long times by taking the stationary phase result for $I''$ and integrating it twice. The terms $I'(t_0)$ and $I(t_1)$ that appear as integration constants are hereby treated as fit parameters. One can deal with the problem of shifting timescales in two ways. The first is to simply apply the fits for integration constants using long-time numerical results. For the present work, such long-time results are numerically costly and hence best avoided. We therefore use a second method: determine the form of the next-highest-order contribution to $I''(t)$, and fit to an intermediate timescale of numerically exact results available for $I''(t)$. This has the added advantage that one can directly observe how the fit extrapolates to longer times. Note that in the noisy cases this does not turn out to be necessary, as $I''$ and $I$ can be seen to be good results on similar timescales that are numerically accessible. This is not the case if the agreement of the approximation is much worse for $I$ than for $I''$.

The form of the next-to-leading order contribution to $I''$ is found to be $\mathcal{O}(\frac{1}{t^2})$ and  includes terms that are modulated by sinusoidal functions. The oscillatory frequencies that enter the problem are determined by characteristic energy scales of the problem such as the energy bandwidth. That the next order term is $\mathcal{O}(\frac{1}{t^2})$ can be verified either through asymptotic matching \cite{Bender} or by an analysis of the contribution from singular regions applied to $I''$, similar to what will be discussed in the following subsection. As an example: restricting the integration bounds of $I''$ evaluated for $\vec{J} = (1,1,\sqrt{2})$ to a small region around the poles, for both uniform field and impurity, gives 
$$I_{\pi/4}''\propto -\int^{\delta}_0 q \cos{(2qt)} dq = \frac{1}{4t^2}\left[ 1 - \cos{(2\delta t)} - 2\delta t\sin{(2\delta t)}\right],$$
where as it will be explained in the next subsection, integration is over a small radius $\delta$ around the singular points. As long as the system is gapless, we expect the singular points to contribute a term of the same power regardless of the particular choice of $\vec{J}$. This is specifically due to the functional form of the dispersion near any $E_k = 0$ point in the gapless phase, which is a Dirac point as long as we are not at a critical point between the gapless and gapped phases, and hence has a linear dispersion. Therefore, the contribution from the stationary points to $I''$ is $\mathcal{O}(\frac{1}{t^2})$. Asymptotic matching can be implemented to check that this is indeed the next order after the leading stationary phase result, and that we are not missing a power between $-1$ and $-2$ \cite{Bender}.

\begin{figure}[htp]
    \centering
    \includegraphics[width=13cm]{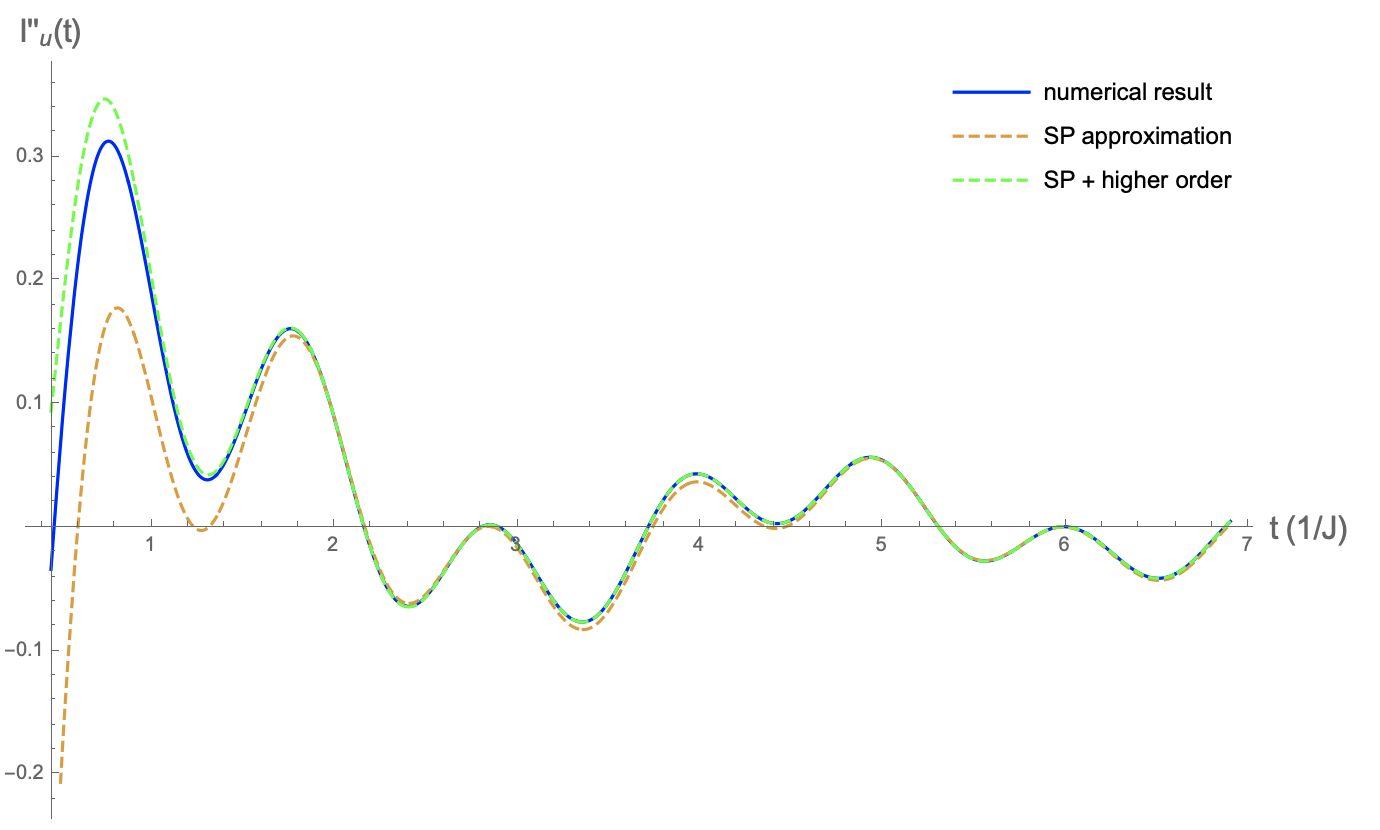}

    \caption{Comparison of numerically exact result for $I_u''$, the first-order asymptotic approximation using stationary phase, and the same approximation augmented with higher-order terms. Here higher-order terms were fitted using numerical results for $I''$ from $t=10$ to $15$. It is noteworthy that the goodness of the fit reaches back to earlier times, and that the asymptotic fit even without higher-order corrections is so good at such relatively short times. \label{fig:IppU}}
\end{figure}

To more accurately capture behavior on numerically accessible timescales, we can therefore use the approximation
$I_{approx}''(t) = I_{sp}''(t) + \frac{1}{t^2}(a + b\sin{(2t)} + c\sin{(6t)} + d\cos{(2t)} + f\cos{(6t)}),$
where $I_{sp}''$ is the result from stationary phase. We fit $a$, $b$, $c$, $d$, and $f$ using a window from the longest-time numerical data that can be calculated. It is noteworthy that augmenting the approximation in this way makes it noticeably better for times earlier than the fitting data, as seen in Fig.\ref{fig:IppU}.

We then compute two indefinite time integrals, at each step fitting the integration constant to the numerical data for $I'$ and $I$. In the end we have an expression for $I_{approx}$ which matches our numerical data well even on intermediate timescales (see Fig.\ref{fig:UniformIntApprox}):

\begin{figure}[htp]
    \centering
    \includegraphics[width=12cm]{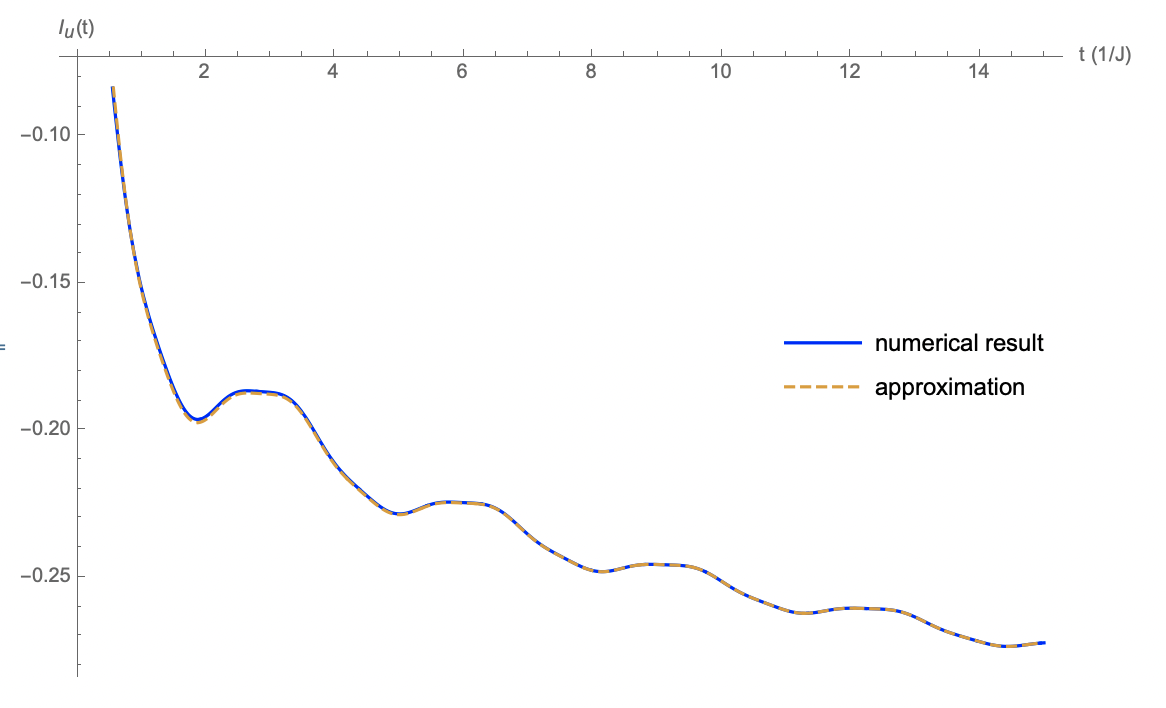}

    \caption{Plot of numerically exact result  and long-time approximation Eq.(\ref{eq:fullUapprox}) for the gapless uniform field integral $I_u$. It is reassuring that while the fit was performed using a window from $t=10$ to $15$, the fit reaches back in time nicely. \label{fig:UniformIntApprox}}
\end{figure}

\begin{align*} \label{eq:fullUapprox}
I(t) &\approx -a + b t + c \cos{(2 t)} - d \cos{(6 t)} + 
e \operatorname{Si}(2 t) \\
&\quad - f t \operatorname{Ci}(2 t) + 
g \operatorname{Ci}(6 t) + h t \operatorname{Ci}(6 t) - 
j \log{t} + k \sin{(2 t)} \\
&\quad - l \sin{(6 t)} - 
m \operatorname{Si}(2 t) + n t \operatorname{Si}(2 t) - 
p \operatorname{Si}(6 t) - q t \operatorname{Si}(6 t),
\end{align*}

\begin{dmath} \label{eq:fullUapprox}
I_u \approx -0.0547401 + 0.215374 t + 0.0000696254 \cos{(2 t)} - 0.0229227 \cos{(6 t)} + 
0.0000696254 \operatorname{Si}(2 t)  - 0.0394651 t \operatorname{Ci}(2 t) + 
0.0000493318 \operatorname{Ci}(6 t) + 0.00115947 t \operatorname{Ci}(6 t) - 
0.0431706 \log{t} + 0.0197325 \sin{(2 t)} - 0.000193245 \sin{(6 t)} - 
0.0598449 \operatorname{Si}(2 t) + 0.000139251 t \operatorname{Si}(2 t) - 
0.000193245 \operatorname{Si}(6 t) - 0.137536 t \operatorname{Si}(6 t),
\end{dmath}
with specific coefficients for $I_u$ as an example given above.  $\operatorname{Si}(z) = \int^z_0 \frac{\sin{t}}{t} dt$ is the sine integral, and $\operatorname{Ci}(x) = \int^{\infty}_x \frac{\cos{(t)}}{t} dt$ is the cosine integral. Finally, to see the leading order behavior we expand for large $t$, finding

$$I \rightarrow c_0 - c_1 t - c_l \log{t}.$$
The linear $c_1$ coefficient in both cases is several orders of magnitude smaller than $c_l$. We note that unlike the other fit parameters it is highly susceptible to small changes in the fit window. This suggests that it might be appearing as a numerical artifact of fit times that are chosen smaller than is ideal. We argue in the next subsection that there are indeed good analytical reasons based on a proper analysis of singularities that tell us that this term should not appear in our asymptotic expansion. Therefore, in the main part of the paper we conclude that the integrals have the asymptotic form reported as: 
\begin{equation}
    I \approx c_0 - c_l \log{t}
\end{equation}
for both the noiseless uniform field and noiseless impurity.

\subsection{Noiseless integrals in the gapless phase: singularity analysis}

In the gapless phase, two integrations of the stationary phase result for the second derivative $I''(t)$ generally gives the following long-time approximation for the integral $I(t)$:
$$I(t) \approx c_0 + c_1 t - c_l \log{t} + \mathcal{O}(1/t),$$
plus some small sinusoidal contributions. Because the stationary phase is applied to the second derivative $I''(t)$, by itself the approximation cannot determine parameters $c_0$ or $c_1$, being that they depend on integration constants. In our case, we chose to fix them numerically by a least square fit. However, we can gain important insight into the value of $c_1$ by thinking about the major asymptotic contributions to $I(t)$ at long times. These contributions are from regions of stationary phase, which will behave like $1/t^2$ as we argued above, and from regions near singularities where $E_k = 0$. Therefore, we expect that in a direct analysis of $I(t)$, the leading order linear and logarithmic behavior must be contributed by regions close to the singularities, rather than from regions of stationary phase. 

In the interest of giving an analytical treatment, we can choose a set of parameters in the gapless phase that leads to a simple form for the integrand around the poles \footnote{Owing to the fact that dispersion around gap closings is generally Dirac-like, we expect the functional form of the contribution from these regions not to be sensitive to the particular choice of $J_x$, $J_y$, and $J_z$.}. For example, the case $\vec{J} = (1, 1, \sqrt{2})$ has $E(\frac{\pi}{4},-\frac{\pi}{4})=E(-\frac{\pi}{4},\frac{\pi}{4}) = 0$. Near these gap closings, the energy takes the form $E(\vec{q}) = 2\sqrt{q_x^2 + q_y^2} \equiv 2q$ where $\vec{q} = \vec{k} - \vec{k}_0$ with $\vec{k}_0$ the stationary point. The contribution from these singularities to the integral in Eq.(\ref{fullimpecho}) for the impurity system has the form
\begin{equation} \label{eq:singularity}
    I_{\pi/4}(t) = 2\int^{\delta}_0 \int^{2\pi}_0 \frac{dq d\theta}{(2\pi)^2} q\frac{\cos{(2qt)-1}}{4q^2} = \frac{1}{\pi} \int^{\delta}_0
dq \frac{\cos{(2qt)}-1}{4q},\end{equation}
where integration is taken over a small circular region of radius $\delta$ around the singular point. This integral gives no linear contribution, with 

\begin{equation} \label{eq:nolin}
I_{\pi/4}(t) \propto -\frac{1}{4\delta^2t^2} - \frac{5\delta^2 t^2}{6} + \frac{1}{2}(3 - 2 \gamma_e - \log{(4\delta^2)} -2\log{t}),
\end{equation}
where $\gamma_e$ is Euler's constant. The lack of a linear term gives us confidence that there is no linear contribution to the integral and therefore no exponential decay in the Loschmidt echo for the noiseless impurity. What's more, the same argument can be applied to the noiseless uniform field, as the contribution to the integral from the singular regions is proportional to Eq.(\ref{eq:singularity}).

One needs to be careful in interpreting Eq.(\ref{eq:nolin}) too naively. Strictly speaking, $\delta$ is a small parameter and $t$ a large one - we must therefore be careful in interpreting terms with the same power in each parameter. One strategy is to think on intermediately long timescales - ones in which $t$ is large but finite, and to think of $\delta$ as vanishingly small. In this case we see that the $\frac{1}{t^2}$ term is likely to be a higher-order correction, but the quadratic term goes to zero and we do not expect it to appear in the asymptotic form for $I(t)$, in agreement with our treatment in the previous sections. In any case, Eq.\eqref{eq:nolin} should be understood not as an exact calculation of the contribution from the singular points, but rather as a heuristic argument reinforcing our interpretation of the linear terms as vanishing, as observed in our augmented stationary phase approach.

\subsection{Noisy integrals} \label{app:noise}

The integrals appearing in Eq.\eqref{Eq:localfid}) and Eq.\eqref{Eq:ufid} require the sort of approximation procedure outlined above, regardless of whether the spectrum is gapped or gapless. This is because of the presence of integrands with the quantity $(E_k - E_{k'})^2$ in the denominator - singularities in the 4D $k$, $k'$ space exist in both cases, and make formally similar contributions to the long-time behavior. This is the origin of the universal form $F(t) \approx C t^{-\alpha} e^{-\beta t}$ for all cases studied; the leading order behavior must be contributed by the singular points, and these contributions are of the same form whether the system is gapped or not. In all noisy cases, integrals displaying singularities turn out to be captured well without introducing higher-order corrections to the stationary phase analysis of $I''$. In this sense their study is simpler than in the noiseless cases, because $I_{approx}''$ can be obtained via stationary phase, and simply integrated twice, at each step fitting the integration constant to the numerical result for $I'$ and $I$.

In both the impurity and uniform field case, a leading-order linear (this corresponds to exponential decay in the Uhlmann fidelity) term is contributed by the first cumulant. In the impurity case this contribution is just $-\kappa t$, and for the uniform field it is given by $-4\kappa h^2 A t C_I$, where $C_I=\int \frac{d^2k}{(2\pi)^2} |u_k|^2$ is a numerical constant that can be computed exactly. Linear terms also arise from the second cumulant in each case, but these terms are $\mathcal{O}(\kappa^2)$ and are higher-order corrections to the linear term from the first cumulant.  For the impurity, the second cumulant gives an integral 
\begin{equation}
    I_{l} = \int \frac{d^2k}{(2\pi)^2} \frac{d^2k'}{(2\pi)^2} \frac{1 - \cos{((E_k - E_{k'})t)}}{(E_k - E_{k'})^2}.
\end{equation}
Two time derivatives gives
$$I_l'' =  \int \frac{d^2k}{(2\pi)^2} \frac{d^2k'}{(2\pi)^2} \cos{((E_k - E_{k'})t)} = \left[ \int \frac{d^2k}{(2\pi)^2} \cos{(E_kt)}\right]^2 + \left[ \int \frac{d^2k}{(2\pi)^2} \sin{(E_kt)}\right]^2 = \left|\int \frac{d^2k}{(2\pi)^2} e^{iE_kt}\right|^2,$$
where we reduce a 2D integral using a trigonometric identity, which is then amenable to a stationary phase approximation using Eq.(\ref{spapp}).

In both gapped and gapless cases we find the asymptotic behavior to be given by 
$$I_l \approx -\gamma + \alpha \log{t} -\beta t,$$
where $\gamma$ can contain some small sinuisoidal pieces. The coefficients found in each case are summarized in TAB.\ref{Tab:impres}.

\begin{table*}[t]
  \centering
\begin{tabular}{|c | c | c | c |} 
\hline 
 $F_l(t) \approx e^{-\kappa t}\exp{[\frac{\kappa^2}{2}(-\gamma + \alpha t -\beta \log{t})]}$  & $\alpha$ & $\beta$ & $\gamma$\\ [1ex]
 \hline
 Gapped & 0.890022 &  0.139317 & 0.447062 \\ [1ex]
 \hline
 Gapless  & 0.644633 & 0.0759909  & 0.274748\\
 [1ex] 
 \hline
\end{tabular}

\caption{Asymptotic behavior ($t\to\infty$) of $F_l(t) \approx e^{-\kappa t}\exp{\left[\frac{\kappa^2}{2}(-\gamma + \alpha t -\beta \log{t})\right]}$ for the noisy impurity. Behavior is formally the same for the gapped and gapless system, but numerical values of coeffecients differ. In actuality $\gamma$ has some small sinusoidal dependence an order of magnitude smaller than its average value, which is given in the table. \label{Tab:impres}}
\end{table*}

The appearance of a positive linear contribution in the second cumulant places a specific upper bound on $\kappa$, which must obviously be small for a perturbative expansion to make sense. The condition is $\kappa > \alpha \frac{\kappa^2}{2}$, or $\kappa < \frac{2}{\alpha}$.

The noisy uniform field case involves more integrals to be approximated, but the methods for calculating them are the same as we have discussed so far. Again, the first cumulant gives the leading-order linear behavior leading to exponential decay, while the contribution from the second cumulant is given by 

\begin{dmath} 
    I_u(t) = 16\kappa^2 h^4 \times \left[ \frac{1}{2}t^2 A\int{ \frac{d^2k}{(2\pi)^2} |v_k|^2|u_k|^2}  - A^2\int{\int \frac{d^2k}{(2\pi)^2} \frac{d^2k'}{(2\pi)^2} |u_k|^2|u_{k'}|^2 \frac{1 - \cos{(E_k + E_{k'})t}}{(E_k + E_{k'})^2}} + A\int{ \frac{d^2k}{(2\pi)^2}|u_k|^2\frac{1 - \cos{2E_kt}}{(2E_k)^2}} + A^2 \int{\int \frac{d^2k}{(2\pi)^2} \frac{d^2k'}{(2\pi)^2} |u_k|^2|u_{k'}|^2 \frac{1 - \cos{(E_k - E_{k'})t}}{(E_k - E_{k'})^2}} \right].
\end{dmath}
As discussed in the main text, we consider large system sizes, for which the $\mathcal{O}(A^2)$ terms dominate. However, the $\mathcal{O}(A)$ terms are not difficult to compute: the first can be computed exactly and gives a small but positive quadratic contribution, signalling that the cumulant expansion will breakdown on some timescale, roughly where $t > \frac{1}{2\kappa h^2}$. The other $\mathcal{O}(A)$ integral, 
$$A\int{ \frac{d^2k}{(2\pi)^2}|u_k|^2\frac{1 - \cos{2E_kt}}{(2E_k)^2}},$$
can be approximated via stationary phase for the gapped system. The contribution that does not decay like $1/t$ comes from the piece not dependent on time, $\int{ \frac{d^2k}{(2\pi)^2}|u_k|^2\frac{1}{(2E_k)^2}}$. Therefore the integral asymptotically converges to the value of that integral. For the gapless system the integral must be treated using the techniques discussed earlier in this appendix. 

The most important behavior comes from the $\mathcal{O}(A^2)$ integrals. The first, 

$$A^2\int{\int \frac{d^2k}{(2\pi)^2} \frac{d^2k'}{(2\pi)^2} |u_k|^2|u_{k'}|^2 \frac{1 - \cos{(E_k + E_{k'})t}}{(E_k + E_{k'})^2}},$$
can be treated directly using a stationary phase approximation for the gapped case. As discussed above, it converges to a constant at long times, given by the value of $A^2\int{\int \frac{d^2k}{(2\pi)^2} \frac{d^2k'}{(2\pi)^2} |u_k|^2|u_{k'}|^2 \frac{1}{(E_k + E_{k'})^2}}$. Interestingly (and conveniently) the same is true even for the gapless system: at long times the integral asymptotically converges to a constant that is given by the time-independent piece. This is likely due to the fact that the presence of singularities is greatly diminished by the structure of the denominator $(E_k +E_{k'})^2$, with singular points only existing in a smaller region of 4D $k$-space, where $E_k = E_{k'} = 0$. Therefore, no interesting time dependence comes from this integral.

Just as in the impurity case, the interesting time dependence is contributed by the integral with $(E_k - E_{k'})^{-2}$ in the integrand. This is treated using the same approach as in the impurity case, further stressing the universality of the formal results, regardless of the presence or lack of a gap or whether the noise is local or global. The results for the noisy uniform field are given in Tab.\ref{Tab:ures}. 

\begin{table*}[t]
  \centering
\begin{tabular}{|c | c | c | c | c |} 
\hline 
 $F_l(t) \approx e^{- \delta A h^2 \kappa t}\exp{[16 A^2 h^4 \kappa^2(-\gamma + \alpha t -\beta \log{t})]}$ & $\delta $ & $\alpha$ & $\beta$ & $\gamma$\\ [1ex]
 \hline
 Gapped & 3.88056  & 0.613024 & 0.139317 & 0.173303\\ [1ex]
 \hline
 Gapless & 3.04973 & 0.366201 & 0.044328 & 0.100994\\
 [1ex] 
 \hline
\end{tabular}

\caption{Asymptotic behavior ($t\to\infty$) of $F_u(t) \approx e^{-\delta A h^2 \kappa t}\exp{\left[16 A^2 h^4 \kappa^2(-\gamma + \alpha t -\beta \log{t})\right]}$ for the noisy uniform field. Behavior is formally the same for the gapped and gapless system, but numerical values of coeffecients differ. The values of $\alpha$ and $\beta$ are decided entirely by the integral $\int{\int \frac{d^2k}{(2\pi)^2} \frac{d^2k'}{(2\pi)^2} |u_k|^2|u_{k'}|^2 \frac{1 - \cos{(E_k - E_{k'})t}}{(E_k - E_{k'})^2}}$, which contributes the dominant asymptotic behavior. \label{Tab:ures}}
\end{table*}

\end{document}